\documentclass[auth]{jalc}

\usepackage{float}
\usepackage{graphicx}
\usepackage{amssymb}
\usepackage{amsmath}
\usepackage{mathtools}
\usepackage{svg}
\usepackage{algorithm}
\usepackage[noend]{algpseudocode}
\usepackage{booktabs}
\usepackage{tikz}
\usepackage{tikz-cd}
\usetikzlibrary{automata, fit, positioning, matrix, arrows.meta, decorations.pathmorphing, shapes, shapes.multipart, chains,shadows.blur}

\usepackage{tikz-qtree}

\usepackage{tikz-dependency}

\tikzset{
-{Latex[length=1.8mm, width=1.4mm]}, 
every initial by arrow/.append style={anchor/.append style={shape=coordinate}},
every state/.style={thick, fill=white, minimum size=8mm},
every loop/.append style=-{Latex[length=1.8mm, width=1.4mm]},
initial text=,
initial distance=0.5cm,
line width=0.7pt,
node distance=18mm,
minimum size=4mm,
}

\definecolor{silver}{rgb}{0.75, 0.75, 0.75}
\definecolor{slategray}{rgb}{0.44, 0.5, 0.56}

\newcommand{\mymacro}[1]{{#1}}
\newcommand{\fapprox}[1]{\widehat{#1}}

\newcommand{\defn}[1]{\emph{#1}}

\usepackage{todonotes}
\definecolor{mintgreen}{RGB}{152, 255, 152}
\makeatletter
\newcommand*\iftodonotes{\if@todonotes@disabled\expandafter\@secondoftwo\else\expandafter\@firstoftwo\fi}  %
\makeatother

\usepackage{colortbl}

\definecolor{ETHBlue}{RGB}{33,92,175}	
\definecolor{ETHGreen}{RGB}{98,115,19}		
\definecolor{ETHPurpleDark}{RGB}{140,10,89}	
\definecolor{ETHPurple}{RGB}{163,7,116}	
\definecolor{ETHGray}{RGB}{111,111,111}	
\definecolor{ETHRed}{RGB}{183,53,45}	
\definecolor{ETHPetrol}{RGB}{0,120,148}	
\definecolor{ETHBronze}{RGB}{142,103,19}	

\colorlet{ETHdarkblue}{ETHBlue!80!black}
\colorlet{ETHdarkgreen}{ETHGreen!80!black}
\colorlet{ETHpink}{ETHPurple}
\colorlet{ETHgray}{ETHGray}
\colorlet{ETHred}{ETHRed}
\colorlet{ETHgreenblue}{ETHPetrol}
\colorlet{ETHbrown}{ETHBronze}

\colorlet{MacroColor}{ETHPetrol}
\colorlet{MACROCOLOR}{MacroColor}

{\itshape}{}
\newtheorem{nclaim1}{Claim}{\itshape}{}
\newtheorem{nclaim2}{Claim}{\itshape}{}
\newtheorem{nclaim3}{Claim}{\itshape}{}

\usepackage{algorithm}
\usepackage[noend]{algpseudocode}  
\algrenewcommand\algorithmicindent{1.0em}%
\newcommand{\rightcomment}[1]{{\color{gray} \(\triangleright\){\footnotesize\textit{#1}}}}
\algrenewcommand{\algorithmiccomment}[1]{\hfill \rightcomment{#1}}  
\algnewcommand{\LinesComment}[1]{\State {\color{black!50!green}\rightcomment{\parbox[t]{.95\linewidth-\leftmargin-\widthof{\(\triangleright\) }}{#1}}}}
\algnewcommand{\LineComment}[1]{\State {\color{black!50!green}\smaller \(\triangleright\) \parbox[t]{\linewidth-\leftmargin-\widthof{\(\triangleright\) }}{\it #1}\smallskip}} 
\algnewcommand{\InlineComment}[1]{\hfill {\color{black!50!green}\(\triangleright\) {\scriptsize \it #1}}}
\algrenewcommand\algorithmicindent{1.0em}%
\algrenewcommand\alglinenumber[1]{{\tiny\color{black!50}#1.}\hspace{-2pt}}
\newcommand{\algorithmicfunc}[1]{\textbf{def} {#1}:}
\algdef{SE}[FUNC]{Func}{EndFunc}[1]{\algorithmicfunc{#1}}{}
\makeatletter
\ifthenelse{\equal{\ALG@noend}{t}}%
  {\algtext*{EndFunc}}
  {}%
\makeatother

\newcommand{\range}[1]{{\mymacro{\mathcal{R}\!\left(#1\right)}}}

\newcommand{\field}{{\mymacro{ \mathbb{F}}}}

\newcommand{\B}{{\mymacro{ \mathbb{B}}}}

\newcommand{\norm}[1]{{\mymacro{ \left\lVert #1 \right\rVert}}}

\newcommand{\pclass}{{\mymacro{\textup{\textsf{P}}}}}
\newcommand{\nlclass}{{\mymacro{\textup{\textsf{NL}}}}}

\newcommand{\pspace}{{\mymacro{\textup{\textsf{PSPACE}}}}}

\newcommand{\monoid}{{\mymacro{\mathbb{M}}}}
\newcommand{\monoidS}{{\mymacro{\mathbb{S}}}}
\newcommand{\transmonoid}[1]{{\mymacro{\mathbb{S}(#1)}}}

\newcommand{\morphismdef}{{\mymacro{\mu}}}
\newcommand{\morphism}[1]{{\mymacro{\mu(#1)}}}

\newcommand{\Boolean}{{\mymacro{\mathbb{B}}}}

\newcommand{\rank}[1]{{\mymacro{\mathrm{rank}(#1)}}}

\newcommand{\alphabet}{{\mymacro{ \Sigma}}}

\newcommand{\kleene}[1]{{\mymacro{#1^*}}}

\def\1{\mathbf{1}}

\def\eps{{\mymacro{ \varepsilon}}}

\newcommand{\str}{{\mymacro{\boldsymbol{w}}}}

\newcommand{\strx}{{\mymacro{\boldsymbol{x}}}}
\newcommand{\stry}{{\mymacro{\boldsymbol{y}}}}
\newcommand{\strz}{{\mymacro{\boldsymbol{z}}}}
\newcommand{\sym}{{\mymacro{w}}}

\newcommand{\syma}{{\mymacro{a}}}
\newcommand{\symb}{{\mymacro{b}}}
\newcommand{\symc}{{\mymacro{c}}}

\newcommand{\set}[1]{{\mymacro{\left\{ #1 \right\}}}}
\newcommand{\powerset}[1]{{\mymacro{\mathcal{P}{\left(#1\right)}}}}
\newcommand{\setcomplement}[1]{{\mymacro{#1^{\textsf{c}}}}}

\newcommand{\N}{{\mymacro{ \mathbb{N}}}}

\newcommand{\nstates}{{\mymacro{ |\states|}}}

\newcommand{\mfstates}{{\mymacro{ \states_{\textsc{mf}}}}}

\newcommand{\nsymbols}{{\mymacro{ |\alphabet|}}}

\newcommand{\graph}{\mymacro{ \mathcal{G}}}

\newcommand{\automaton}{{\mymacro{ \mathcal{A}}}}

\newcommand{\detautomaton}{{\mymacro{ \automaton_{\textsc{det}}}}}
\newcommand{\detautomatonprime}{{\mymacro{ {\automaton}'_{\textsc{det}}}}}
\newcommand{\mooreautomatonprime}{{\mymacro{ {\automaton}'_{\textsc{moore}}}}}
\newcommand{\mfautomatonprime}{{\mymacro{ {\automaton}'_{\textsc{mf}}}}}
\newcommand{\ssautomaton}{{\mymacro{ \automaton_{\textsc{ss}}}}}
\newcommand{\powerautomaton}{{\mymacro{ \automaton_{\textsc{pow}}}}}
\newcommand{\ssautomatonprime}{{\mymacro{ \automaton'_{\textsc{ss}}}}}
\newcommand{\ssstates}{{\mymacro{ \states_{\textsc{ss}}}}}
\newcommand{\sstrans}{{\mymacro{ \trans_{\textsc{ss}}}}}
\newcommand{\ssfinal}{{\mymacro{ \final_{\textsc{ss}}}}}

\newcommand{\stateq}{{\mymacro{ q}}}

\newcommand{\states}{{\mymacro{ Q}}}

\newcommand{\trans}{{\mymacro{ \delta}}}

\newcommand{\apath}{{\mymacro{ \boldsymbol \pi}}}

\newcommand{\initial}{{\mymacro{ I}}}

\newcommand{\final}{{\mymacro{ F}}}

\newcommand{\languagefsa}[1]{{\mymacro{\mathcal{L} ( #1 )}}}

\newcommand{\mooreautomaton}{{\mymacro{ \automaton_{\textsc{moore}}}}}
\newcommand{\mfautomaton}{{\mymacro{ \automaton_{\textsc{mf}}}}}
\newcommand{\mmfautomaton}{{\mymacro{ \automaton_{\textsc{mmf}}}}}
\newcommand{\brautomaton}{{\mymacro{ \automaton_{\textsc{br}}}}}
\newcommand{\brautomatonprime}{{\mymacro{ \automaton'_{\textsc{br}}}}}
\newcommand{\universalautomaton}{{\mymacro{ \automaton_{\textsc{univ}}}}}

\newcommand{\fsatuple}{{\mymacro{ \left( \alphabet, \states, \initial, \final, \trans \right)}}}

\newcommand{\reverse}[1]{{\mymacro{ {#1}^R}}}
\newcommand{\edgenoweight}[3]{#1 \xrightarrow{#2} #3}

\newcommand{\powerstate}{{\mymacro{ \mathcal{Q}}}}
\newcommand{\initpowerstate}{{\mymacro{ \mathcal{Q}_{\initial}}}}

\def\sC{{{\mymacro{ \mathcal{C}}}}}

\def\sF{{{\mymacro{ \mathcal{F}}}}}

\def\sI{{{\mymacro{ \mathcal{I}}}}}

\def\sL{{{\mymacro{ \mathcal{L}}}}}

\newcommand{\bigO}[1]{{\mymacro{ \mathcal{O}\left(#1\right)}}}

\AtEndPreamble{%
  \usepackage{cleveref}
}

\AddToHook{begindocument/before}{
  \crefformat{section}{Section #2#1#3}
  \Crefformat{section}{Section #2#1#3}
  \crefrangeformat{section}{Sections #3#1#4--#5#2#6}
  \crefmultiformat{section}{Section #2#1#3}{ and Section #2#1#3}{, Section #2#1#3}{, and~Section #2#1#3}
  \crefformat{subsection}{Section #2#1#3}
  \Crefformat{subsection}{Section #2#1#3}
  \crefrangeformat{subsection}{Sections #3#1#4--#5#2#6}
  \crefmultiformat{subsection}{Section #2#1#3}{ and~Section #2#1#3}{, Section #2#1#3}{, and~Section #2#1#3}
  \crefname{table}{Tab.}{Tabs.}
  \crefname{figure}{Figure}{Figures}
  \crefname{algorithm}{Algorithm}{Algorithms}
  \crefname{equation}{Eq.}{Eqs.}
  \crefname{example}{Example}{Examples}
  \crefname{fact}{Fact}{Facts}
  \crefname{appendix}{Appendix}{Appendices}
  \crefname{theorem}{Theorem}{Theorems}
  \crefname{reTheorem}{Theorem}{Theorems}
  \crefname{aquestion}{Question}{Questions}
  \crefname{assumption}{Assumption}{Assumptions}
  \crefname{lemma}{Lemma}{Lemmas}
  \crefname{reLemma}{Lemma}{Lemmas}
  \crefname{proposition}{Proposition}{Propositions}
  \crefname{chapter}{Chapter}{Chapters}
  \crefname{line}{line}{lines}
  \crefname{principle}{Principle}{Principles}
  \crefname{definition}{Definition}{Definitions}
  \crefname{corollary}{Corollary}{Corollaries}
  \crefname{Exercise}{Exercise}{Exercises}
  \crefname{observation}{Observation}{Observations}
}

\begin{document}%
     %
\title{Towards A Parametrized Complexity of Determinization}
     %

\runningtitle{Towards A Parametrized Complexity of Determinization}
     
\runningauthors{\textsc{I. Baburin}, \textsc{R. Cotterell}}
     

\author[VUW]{Ivan Baburin\orcidID{0009-0008-2647-5836}} 

\author[ETH]{Ryan Cotterell\orcidID{0000-0003-4080-1833}} 
\address[VUW]{School of Mathematics and Statistics,\\
  Victoria University of Wellington\\
  Gate 7, Kelburn Parade, Wellington 6012, New Zealand\\
  \email{ivan.baburin@alumni.ethz.ch}
}
\address[ETH]{Department of Computer Science,\\
  ETH Zurich\\
  Andreasstrasse 5, 8092 Zürich, Switzerland\\
  \email{ryan.cotterell@inf.ethz.ch}
}



\maketitle
%
\begin{abstract}
  Given a nondeterministic finite automaton (NFA), we want to estimate the difficulty of finding an equivalent deterministic finite automaton (DFA).
  We demonstrate that computing the state complexity of an NFA within polynomial precision is $\pspace$-hard, and furthermore the same holds for the complexity of the subset construction, Brzozowski's algorithm, or, indeed, \emph{any} determinization algorithm which runs at least as fast.
  This result implies that making any \textit{a priori} estimate of the running time of the major determinization algorithms is inherently hard. To circumvent this---in the spirit of parametrized complexity---we try to identify complexity measures which can guarantee efficient determinization. Our main idea rests on linking the complexity of the subset construction to transition monoids using the semigroup complexity. We show that the semigroup complexity by itself can be reasonably approximated for certain classes of NFAs, and then introduce \emph{subset complexity}---a refinement of the semigroup complexity which has a tighter connection with the subset construction. In certain scenarios, the subset complexity can be approximated in polynomial time, making it potentially viable for more practical applications.
\end{abstract}

%
%
%

\section{Introduction}\label{sec:introduction}
The equivalence in expressivity between nondeterministic finite automata (NFAs) and deterministic finite automata (DFAs) was one of the first foundational results in the theory of finite automata. The original proof by Rabin and Scott~\cite{RabinPowerset} introduced the powerset construction, a general algorithm for converting an arbitrary NFA into an equivalent DFA.
Despite its simplicity and pedagogical value, a major drawback of the powerset construction is that it yields a DFA with exactly $2^n$ states, where $n$ is the number of states in the original NFA. Unfortunately, the equivalent DFA produced by the powerset construction is in general \emph{not} minimal and can, thus, often be reduced using standard finite minimization techniques~\cite{automata-book}. Moreover, in certain pathological cases, the minimal DFA $\detautomaton$ is \emph{exponentially} smaller than the DFA $\powerautomaton$ returned by the powerset construction, meaning the algorithm may perform an exponential amount of unnecessary output. A trivial example illustrates this phenomenon: consider an arbitrary $n$-state NFA with \emph{no} final states. In this case, the minimal DFA is always the $1$-state automaton accepting the empty language. Furthermore, this exponential blow-up is also unavoidable in the worst case---Moore~\cite{moore1971bounds} demonstrated the existence of an $n$-state NFA, known as Moore's automaton, for which the minimal equivalent DFA requires \emph{exactly} $2^n$ states, i.e., the exponentially large DFA produced by the powerset construction is already minimal. Subsequent research identified even simpler classes of NFAs for which the powerset construction yields the minimal DFA~\cite{BORDIHN20093209}.

Besides the powerset construction, several alternative determinization algorithms have been proposed, such as the classic subset construction \cite{allauzen2007openfst}, Brzozowski's algorithm~\cite{brzozowski}, width-based determinization \cite{width-NFA} and many others \cite{five-determinization}. However, there has been only little work analyzing the actual performance of determinization algorithms: some theoretical analysis of the complexity of the subset construction on homogeneous automata has been presented in \cite{subsethomogenous}, the average-case complexity of Brzozowski's algorithm was analyzed in \cite{brzozowski-performance}, and some more empirical approaches were used in \cite{five-determinization}. This lack of attention can be explained by the seeming detachment of determinization complexity from any classical complexity metrics defined on the NFAs,\footnote{See the discussion in \cite{complexity-survey}.} such as, for example, the state complexity $|\detautomaton|$: it is known that even NFAs with a state complexity of exactly $1$ can be notoriously difficult to determinize \cite{pspace-survey}. Moreover, depending on the choice of the determinization algorithm for a specific NFA, the performance could vary significantly, hence it is difficult to come up with a general measure for it. 

Despite these shortcomings, determinization remains one of the key transformations for finite automata, since many language operations such as complement, equivalence, and inclusion testing can be realized on DFAs in polynomial time, while in general, have worst-case exponential time complexity on NFAs \cite{operations-complexity-nfas}.
As we see it, a complete theory for finite automata ought to answer the following questions: when can the transformation of an NFA into an equivalent DFA be done in polynomial time, and what is the best algorithm to achieve that? A satisfactory answer to this question could be an analog of \emph{parametrized} complexity for NFA determinization. Unfortunately, due to all the difficulties mentioned above, there seems to be no obvious way to approach this problem, which is the primary motivation for this work.

In the first part of the article (\cref{sec:determinization-difficulties,sec:algorithmic-landscape,sec:inapproximability}) we develop a theoretical framework to analyze the complexity of determinization. Specifically, in \cref{sec:determinization-difficulties} we discuss the systematic difficulties behind classifying NFAs with respect to the time complexity of determinization, and show why many classical metrics, such as the state complexity, are destined to fail. In \cref{sec:algorithmic-landscape} we introduce a natural way (denoted with relation $\geq$) to compare algorithms with worst-case exponential time complexities: a determinization algorithm \emph{improves} on another one, if it is at least as fast on every possible input. These ideas culminate in \cref{sec:inapproximability}, where we show that the running time of \emph{any} determinization algorithm \textsc{Alg} which is at least as fast as the subset construction (or at least as fast as Brzozowski's algorithm) is inherently unpredictable, as summarized below:
\begin{center}
   \begin{tikzpicture}
       \node[align=center, font={\normalsize}] at (0,1) {\textsc{Alg}};
       \node[align=center, font={\normalsize}] at (0.75,1) {$\geq$};

       \node[align=center, font={\small}] at (2.1,1) {Subset \\ construction\\ (\cref{thm:inapproximability})};
       \node[align=center, font={\normalsize}] at (3.625,1) {or};

       \node[align=center, font={\normalsize}] at (4.5,1) {\textsc{Alg}};
       \node[align=center, font={\normalsize}] at (5.25,1) {$\geq$};
       \node[align=center, font={\small}] at (6.6,1) {Brzozowski's \\ algorithm \\ (\cref{thm:inapp-brzozowski})};
       \node[align=center, font={\normalsize}] at (8.3,1) {$\Longrightarrow$};
       
       \node[align=center, font={\small}] at (10.25,1) {approximating\\ runtime of \textsc{Alg}\\ is \pspace-hard};
   \end{tikzpicture}
\end{center}
Furthermore, a weaker version of this statement holds even if the exact state complexity of the underlying NFA is known (Corollary \ref{corr:known-state}).

In the second part of the article (\cref{sec:semigroup-approach,sec:subset-complexity,sec:practical}), we focus on the performance of the subset construction, since it remains the backbone of many modern determinization algorithms \cite{five-determinization}. In particular, its time complexity can be naturally parametrized using the size of the produced DFA, the subset automaton $\ssautomaton$. In \cref{sec:semigroup-approach} we discuss the semigroup complexity---a known measure that naturally bounds the size of the subset automaton, and derive explicit bounds for some restricted subclasses of NFAs (specifically unary, permutation, commutative, and dense automata). \cref{sec:subset-complexity} then further develops this approach by introducing the subset complexity: a new measure that refines the semigroup complexity, while still bounding the size of the subset automaton. This creates the following hierarchy of complexity measures for non-deterministic automata:
\begin{center}
   \begin{tikzpicture}
       \node[align=center, font={\small}] at (8.75,1) {state \\ complexity \\ $|\detautomaton|$};
       \node[align=center, font={\normalsize}] at (7.5,1) {$\geq$};
       \node[align=center, font={\small}] at (6.25,1) {subset \\ automaton \\ $|\ssautomaton|$};
       \node[align=center, font={\normalsize}] at (5,1) {$\geq$};
       \node[align=center, font={\small}] at (3.75,1) {subset \\ complexity\\ (\cref{sec:subset-complexity})};
       \node[align=center, font={\normalsize}] at (2.5,1) {$\geq$};
       \node[align=center, font={\small}] at (1.25,1) {semigroup\\ complexity \\ (\cref{sec:semigroup-approach})};
   \end{tikzpicture}
\end{center}
Finally, \cref{sec:practical} discusses computational aspects of the subset complexity, specifically scenarios where it can be bounded by a polynomial---something of relevance for determinization in practice.

This article forms an extended version of results partially presented in \cite{onthefly} and \cite{close-analysis}.

\section{Background and Definitions}\label{sec:background}

We start the technical portion of the paper by briefly reciting necessary definitions concerning finite automata; for a more comprehensive introduction, we refer to \cite{automata-book}.
An \defn{alphabet} $\alphabet$ is a finite, non-empty set of symbols.
With $\kleene{\alphabet}$, we denote the set of all finite strings, also called \defn{words}, over $\alphabet$.
The symbol $\eps \in \kleene{\alphabet}$ denotes the empty word, i.e., the unique word of length 0.
Two words $\stry$ and $\strx$ are placed adjacent to each other, e.g., $\strx \stry$  denotes the concatenation of those words.
\begin{definition}
    A \defn{finite automaton (FA)} $\automaton = \fsatuple$ is a quintuple where $\alphabet$ is an alphabet, $\states$ is a finite set of states, $\initial \subseteq \states$ is a set of the initial states, $\final \subseteq \states$ is a set of accepting states and $\trans \subseteq \states \times (\alphabet \cup \set{\eps}) \times \states$ is a finite multiset of labeled transitions between states.
\end{definition}
Note that a transition labeled with $\eps$ can be performed at any point.
The \defn{size} of a finite automaton $\automaton =  \fsatuple$ is its number of states $|\states|$; we use the cardinality operator $| \automaton |$ to denote the size. 

For the remainder of this article, we assume that all automata are $\eps$-free, i.e., contain no $\eps$ transitions.
This assumption is without loss of generality as there exist efficient procedures for removing $\eps$-transitions \cite{allauzen2007openfst}. 

\begin{definition}\label{def:path}
    A \defn{path} $\apath = \edgenoweight{\stateq_0}{\sym_1}{\stateq_1} \edgenoweight{}{\sym_2}{\ldots} \edgenoweight{}{\sym_{N}}{\stateq_N}$ in an FA $\automaton$ is a labeled sequence of states, where each label $\sym_n \in \alphabet \cup \{\eps\}$ is a symbol from the alphabet, and each $\stateq_n \in \states$ is a state in $\automaton$, such that for all $n$ we have $(\stateq_{n-1}, \sym_n, \stateq_{n}) \in \trans$.
    We will refer to the concatenation of symbols along the path $\sym_1 \cdots \sym_{N}$ as the path's \defn{yield}. 
\end{definition}

We say that a word $\str \in \kleene{\alphabet}$ is \defn{recognized} by the FA $\automaton$ if there exists a path in $\automaton$ from some initial state $\stateq_\initial \in \initial$ to some final state $\stateq_F \in \final$ with yield $\str$. 
Moreover, we denote with $\languagefsa{\automaton} \subseteq \kleene{\alphabet}$ the \defn{language} (set of all words) recognized by $\automaton$. 
Two FAs $\automaton$ and $\automaton'$ are called \defn{equivalent} if they accept the same language, i.e., if $\languagefsa{\automaton} = \languagefsa{\automaton'}$. Given two languages $\sL$ and $\sL'$ we define their \defn{concatenation}
\begin{equation}
    \sL\sL' = \{ \strx \stry  \ | \ \strx \in \sL, \ \stry \in \sL' \}
\end{equation}
as the language accepting all concatenations of word pairs in them. In case $\sL' = \{ \str \}$ is a singleton, we overload the notation and simply write $\sL\str$ for brevity.

\begin{definition}\label{def:trim}
A state $\stateq \in \states$ is called \defn{accessible} if there exists a path from $\initial$ to $\stateq$, and \defn{co-accessible} if there exists a path from $\stateq$ to $\final$. An automaton where all states are both accessible and co-accessible is called \defn{trim}.
\end{definition}

Similarly to $\eps$-transitions, we assume that all automata we deal with are trim, since removing inaccessible states can be trivially done using a depth-first search in linear time.
\begin{definition}\label{def:reverse}
    The \defn{reverse} FA of an automaton $\automaton$, marked by $\reverse{\automaton}$, is an FA obtained by reversing the direction of all transitions in $\automaton$ and swapping the initial and final states. 
\end{definition}

We denote the subroutine transforming any FA $\automaton$ into its reverse $\reverse{\automaton}$ as \textsc{Reverse}, and note that $ \languagefsa{\reverse{\automaton}} = \reverse{\languagefsa{\automaton}}$, i.e., the set of all reversed words from $\languagefsa{\automaton}$. 

\begin{definition}\label{def:deterministic}
    An FA \textup{$\automaton = \fsatuple$} is \defn{deterministic} if and only if the initial state is unique, i.e., $|\initial| = 1$, and for every $(\stateq, \sym) \in \states \times \alphabet$ there is exactly one $\stateq' \in \states$ such that $(\stateq, \sym, \stateq') \in \trans$, i.e., the relation corresponds to a total function of type $\states \times \alphabet \rightarrow \states$.
    An FA that is not deterministic is called \defn{nondeterministic} and an FA whose reverse $\reverse{\automaton}$ is deterministic is called \defn{co-deterministic}.
\end{definition}

As a shorthand, for any $\str \in \kleene{\alphabet}$, we write $\trans(\stateq, \str) = \trans(\cdots \trans( \trans(\stateq, \sym_1), \sym_2) \cdots, \sym_N)$ to denote an iterative application of the transition function to the word $\str= \sym_1 \cdots \sym_N$. Typically, DFAs are both easier to implement and to use---many properties, e.g., language equivalence between FAs, are hard to decide for general NFAs, but can be decided for deterministic automata in polynomial time \cite{pspace-survey}. 

\subsection{Powerset Construction}\label{sec:background-powerset}

It is well-known that NFAs and DFAs recognize the same class of languages: this fact is typically proven using the \emph{powerset construction}. We define a naive procedure \textsc{PowerConstr} which, given an arbitrary NFA $\automaton = \fsatuple$, constructs an equivalent DFA $\powerautomaton = (\alphabet, 2^\states, \powerstate_\initial, \final_\textsc{pow}, \trans_\textsc{pow})$, known as the \emph{powerset automaton}. Every state $\powerstate \in 2^\states$, also called the \emph{powerstate}, satisfies $ \powerstate \subseteq \states$ and all transitions between powerstates in $\trans_\textsc{pow}$ satisfy the condition
\begin{equation}
    \edgenoweight{\powerstate_i}{\sym}{\powerstate_j} \quad \Longleftrightarrow \quad \exists \stateq_i \in \powerstate_i \  \exists \stateq_j \in \powerstate_j \quad \text{such that} \quad (\stateq_i, \sym, \stateq_j) \in \trans.
\end{equation}
Furthermore, $\powerstate_\initial = \initial$ is a powerstate containing all initial states of $\automaton$ and $\final_\textsc{pow}$ is a set of powerstates containing at least one state from the final states $\final$. By construction, $\powerautomaton$ is a DFA with $|\powerautomaton| = 2^{|\automaton|}$ and one can easily show that $\languagefsa{\powerautomaton} = \languagefsa{\automaton}$.

More generally, transforming a nondeterministic automaton into an equivalent deterministic one is commonly referred to as \defn{determinization}. Because the resulting DFA is not unique in general, we will slightly narrow down this definition and require valid determinization algorithms to produce the \defn{minimal} equivalent DFA $\detautomaton$ for each NFA $\automaton$ (sometimes this is referred to as \emph{canonization}, and $\detautomaton$ is always unique up to isomorphism \cite{automata-book}). Overall, this has little effect on the underlying complexity, since DFA minimization can be performed in quasi-linear time by means of Hopcroft's minimization algorithm \cite{hopcroft-minimization}. In particular, we can easily apply Hopcroft's minimization algorithm (denote it by \textsc{Minimize}) to the output DFA $\powerautomaton$ from \textsc{PowerConstr}, which yields a valid determinization algorithm:
\begin{equation}
    \detautomaton = \textsc{Minimize}(\textsc{PowerConstr}(\automaton)). 
\end{equation}
\begin{definition}\label{def:state-complexity}
    Given an FA $\automaton$, we call the number of states in its minimal equivalent deterministic automaton $\detautomaton$ its \defn{state complexity}, usually denoted as $|\detautomaton|$.
\end{definition}

\subsection{Subset Construction}\label{sec:background-subset}

In practice, the most commonly implemented building block for determinization algorithms is the so-called \defn{subset construction} \cite{allauzen2007openfst}, a dynamic version of the powerset construction which only keeps track of accessible states from $\powerautomaton$; hence, the states and transitions in the DFA returned by the subset construction are always a subset of the states and transitions of $\powerautomaton$. \cref{alg:subsetconstruction} provides the pseudocode for an example implementation of the subset construction and, analogously to the powerset construction, it can be coupled with \textsc{Minimize} to obtain a valid determinization algorithm:
\begin{equation}\label{eq:subset-constr}
        \detautomaton = \textsc{Minimize}(\textsc{SubsetConstr}(\automaton)).
\end{equation}

Note that this creates a slight ambiguity---the term subset construction may refer both to \cref{alg:subsetconstruction} and the composed determinization algorithm in \cref{eq:subset-constr}.
To distinguish the two, we will use the term \textsc{SubsetConstr} for \cref{alg:subsetconstruction}, and reserve the term subset construction for the composed algorithm. \textsc{SubsetConstr} roughly works as follows:
given an input NFA $\automaton = \fsatuple$, it explores all subsets of states accessible from the powerstate $\powerstate_\initial$ (composed of initial states $\initial$) by following the labeled transitions from $\trans$. 
Each newly found state $\powerstate \subseteq \states$ will be placed on the stack, and for every $\powerstate$ on the stack, we search for all of its direct neighbors and place them on the stack and then remove $\powerstate$. This process is repeated iteratively until no new neighbors can be found, i.e., the stack becomes empty. Algebraically, this can be seen as multiplying the characteristic vector for the initial state $\powerstate_\initial$ with all possible sequences of transition matrices and terminating as soon as no new states can be produced. 
We will refer to the DFA $\ssautomaton$ obtained by applying \cref{alg:subsetconstruction} on $\automaton$ as the \defn{subset automaton}.
\begin{algorithm}[t]
\caption{(\textsc{SubsetConstr})}\label{alg:subsetconstruction}

\begin{algorithmic}
\Require $\automaton = \fsatuple$ 
\State $\initpowerstate \gets \{ \stateq \ | \ \stateq \in \initial \}$
\State $\ssautomaton \gets (\alphabet, \ssstates \coloneqq \emptyset, \powerstate_{\initial}, \ssfinal \coloneqq \emptyset, \sstrans \coloneqq \emptyset)$
\State \textbf{push} $\powerstate_\initial$ on the $\mathsf{stack}$
\State $\ssstates \gets \{ \powerstate_\initial \}$
\While{$|\mathsf{stack}| > 0$}
    \State \textbf{pop} $\powerstate$ from the $\mathsf{stack}$
    \ForAll{$\sym \in \alphabet $}
        \State $\powerstate' \gets \{\stateq' \ | \ (\stateq, \sym, \stateq') \in \trans, \ \stateq \in \powerstate \}$
        \State $\sstrans \gets \sstrans \cup \{\edgenoweight{\powerstate}{\sym}{\powerstate'} \}$
        \If{$\powerstate' \notin \ssstates$}
            \State $\ssstates \gets \ssstates \cup \{ \powerstate' \}$
            \State \textbf{push} $\powerstate'$ on the $\mathsf{stack}$
        \EndIf
    \EndFor
\EndWhile
\State $\ssfinal \gets \{ \powerstate \in \ssstates \ | \ \powerstate \cap \final \neq \emptyset \}$
\State \textbf{return} $\ssautomaton$
\end{algorithmic}
\end{algorithm}

\subsection{Brzozowski's Algorithm}\label{sec:background-brzozowski}

\cref{alg:subsetconstruction} plays a key role in automata determinization far beyond the classical subset construction, as illustrated in the next result due to Brzozowski \cite{brzozowski}.
\begin{theorem}[(Brzozowski \cite{brzozowski})]\label{thm:brzozowski}
    If an NFA $\automaton$ is trim and co-deterministic, then \cref{alg:subsetconstruction} produces the minimal DFA \textup{$\detautomaton$}.
\end{theorem}

\cref{thm:brzozowski} gave birth to a novel determinization algorithm based on \cref{alg:subsetconstruction}, which directly produces the minimal DFA given any trim NFA $\automaton$ as input. This algorithm is known as \defn{Brzozowski's algorithm}, defined as 
\begin{equation}
    \detautomaton = \textsc{SubsetConstr}(\textsc{Reverse}(\textsc{SubsetConstr}(\textsc{Reverse}(\automaton)))).   
\end{equation}
Several other determinization algorithms based on \cref{alg:subsetconstruction} exist, some examples can be found in \cite{five-determinization}. Herewith, we wrap up our discussion of the algorithmic aspects of automata and move on to an alternative way of characterizing FAs through the algebraic approach using semigroups defined by the underlying transition matrices.

\subsection{Transition Monoids}

\begin{definition}
    Consider an FA $\automaton = \fsatuple$ and, for each $\sym \in \alphabet$, define a binary relation ${T}^{(\sym)}$ over $\states$ with $(\stateq, \stateq') \in T^{(\sym)} \Leftrightarrow (\stateq, \sym, \stateq') \in \trans $ for all $\stateq, \stateq' \in \states$. A \defn{transition monoid} $\transmonoid{\automaton}$ is a binary relation monoid over $\states$ generated by $ \{ T^{(\sym)} \ | \ \sym \in \alphabet \}$ and closed under the relation composition operator $\circ$ with the identity relation \textup{$\mathrm{id}$}.
\end{definition}

It is often more convenient to think of $\transmonoid{\automaton}$ as a monoid of $\nstates \times \nstates$ matrices over the Boolean semifield $\Boolean \coloneqq \langle \{ 0, 1 \}, \vee, \wedge, 0, 1 \rangle$ closed under Boolean matrix multiplication with the identity element $\sI$, such that:
\begin{equation}
    T^{(\sym)}_{i, j} = 1 \quad \Longleftrightarrow \quad (\stateq_i, \stateq_j) \in T^{(\sym)}.
\end{equation}
Notice that the monoid $\transmonoid{\automaton}$ is, by construction, closely related to the regular language $\languagefsa{\automaton}$.

\begin{definition}
    A language $\sL \subseteq \kleene{\alphabet}$ is \defn{recognized by a monoid} $\monoid$ of binary relations over $\states$ if there exists a surjective morphism $\morphismdef : \kleene{\alphabet} \rightarrow \monoid$ and a subset  $\monoid_{\final} \subseteq \monoid$, called an \defn{accepting subset}, such that $\sL = \morphismdef^{-1}(\monoid_{\final})$.
\end{definition}

From the definition above it is immediately visible that $\transmonoid{\automaton}$ recognizes the language $\languagefsa{\automaton}$. Indeed, if we view $\initial$ and $\final$ as Boolean vectors we can define the accepting subset as 
\begin{equation}
    \monoidS_{\final} \coloneqq \left\{ M \in \transmonoid{\automaton} \ | \ \initial^\top M \final \neq 0 \right\}
\end{equation}

\noindent and a morphism $\morphismdef$ for an arbitrary word $\str = \sym_1\sym_2\sym_3 \ldots \sym_m \in \kleene{\alphabet}$
\begin{equation}
    \morphism{\str} = \sI \cdot \prod^m_{i=1} T^{(\sym_i)} \in \transmonoid{\automaton}
\end{equation}

\noindent with $\morphism{\eps} = \sI$. As a consequence, the inverse $\morphismdef^{-1}(\monoidS_{\final})$ only contains words $\str \in \kleene{\alphabet}$ that are a yield of at least one path from $\initial$ to $\final$ in $\automaton$.
The duality between automata and semigroups has been extensively studied; for a more detailed overview, we refer to \cite{pin1997syntactic}.

\section{Troubles with Determinization}\label{sec:determinization-difficulties}

As mentioned in \cref{sec:introduction}, the problem of converting an NFA into an equivalent DFA is difficult because the state complexity of many NFAs can be exponentially large. 
Two classical examples demonstrating this phenomenon are the $n$-state Moore automaton \cite{moore1971bounds} and the $n$-state Meyer--Fischer automaton \cite{meyerfischer}, both of which exhibit a state complexity of $2^n$; see \cref{fig:Moore-NFA}. 

\begin{example}
    The blow-up occurring in a Meyer--Fischer automaton $\mfautomaton$ has a particularly intuitive explanation. 
    Let ${| \str |}_{\syma}$ denote the number of times the symbol $\syma$ appears in a word $\str \in  \kleene{\{\syma, \symb \}}$.
    Then, the language accepted by an $n$-state $\mfautomaton$ can be described as
    \begin{equation}
        \languagefsa{\mfautomaton} = \{ \strx \stry  \ | \ \strx \in \{ \varepsilon \} \cup (\kleene{\{\syma, \symb\}}\symb), \; \stry \in \kleene{\{\syma, \symb\}} \text{ and } {|\stry|}_{\syma} \equiv 0 \pmod{n}\}. 
    \end{equation}
    For every word $\str \in  \languagefsa{\mfautomaton}$, the nondeterministic Meyer--Fischer automaton guesses a prefix that ends with $\symb$ (or is empty), such that the number of symbols $\syma$ in the remaining suffix is a multiple of $n$. 
    To accomplish that, any deterministic counterpart of $\mfautomaton$ needs to keep track of all parities modulo $n$ for all occurrences of $\symb$ in $\str$, i.e., exactly $2^n$ possible options.
    \begin{figure}[tb]
    \centering
    \begin{tikzpicture}[->,>=stealth',shorten >=0.5pt,auto,node distance=15mm, scale=0.8, transform shape]
        \tikzstyle{every state}=[fill=none,draw=black,text=black,minimum size=0pt,inner sep=2.5pt]
            \node[state, initial] (q1) {$q_1$};
            \node[state, right of=q1] (q2) {$q_2$};
            \node[state, right of=q2] (q3) {$q_3$};
            \node[fill=white, right of=q3] (q?) {$\ldots$};
            \node[state, accepting, right of=q?] (qn) {$q_n$};
            \draw (q1) edge[loop above] node{$b$} (q1)
                  (q1) edge[above] node{$a$} (q2)
                  (q2) edge[above] node{$a, b$} (q3)
                  (q3) edge[above] node{$a, b$} (q?)
                  (q?) edge[above] node{$a, b$} (qn)
                  (qn) edge[bend right=1.7cm, above] node{$a$} (q2)
                  (qn) edge[bend left=1.5cm, below] node{$a$} (q1);
    \end{tikzpicture}
    \hfill
    \begin{tikzpicture}[->,>=stealth',shorten >=0.5pt,auto,node distance=15mm, scale=0.8, transform shape]
        \tikzstyle{every state}=[fill=none,draw=black,text=black,minimum size=0pt,inner sep=2.5pt]
            \node[state, accepting, initial] (q1) {$q_1$};
            \node[state, above right = {12mm} and {3mm} of q1] (q2) {$q_2$};
            \node[state, above right of=q2] (q3) {$q_3$};
            \node[state, right of=q3] (q4) {$q_4$};
            \node[fill=white, below right of=q4] (q?) {$\ldots$};
            \node[state, right = {43mm} of q1] (qn) {$q_n$};

            \draw (q1) edge[in=85,out=115,loop, above] node{$b$} (q1)
                  (q2) edge[loop above] node{$b$} (q2)
                  (q3) edge[loop above] node{$b$} (q3)
                  (q4) edge[loop above] node{$b$} (q4)
                  (qn) edge[in=50,out=80,loop, above] node{$b$} (qn)
                  (q1) edge[above left] node{$a$} (q2)
                  (q2) edge[above left] node{$a$} (q3)
                  (q3) edge[above] node{$a$} (q4)
                  (q4) edge[above right] node{$a$} (q?)
                  (q?) edge[above right] node{$a$} (qn)
                  (qn) edge[above] node{$a, b$} (q1)
                  (q2) edge[bend left=1cm,left] node{$b$} (q1)
                  (q3) edge[bend left=1cm,left] node{$b$} (q1)
                  (q4) edge[bend left=1cm,above] node{$b$} (q1);
        \end{tikzpicture}
    \caption{An $n$-state Moore automaton $\mooreautomaton$ (left) and an $n$-state Meyer--Fischer automaton $\mfautomaton$ (right). Both of them are nondeterministic with a state complexity of $2^n$ for every $n \geq 2$.}
    \label{fig:Moore-NFA}
    \end{figure}
\end{example}

This is enough to conclude that there cannot exist an efficient (i.e., polynomial-time in the size of the NFA) procedure for converting NFAs into minimal DFAs.
Nevertheless, it is worth emphasizing that not all NFAs will exhibit exponential blow-up, since many classes of NFAs---the most naive example being automata with no accepting states---can be converted into an equivalent DFA quite efficiently \cite{width-NFA}. It is natural to raise the following question: 

\begin{quote}
    \textit{How should one quantify the complexity of NFA determinization?}
\end{quote}

Before we attempt to answer the question, we first need to formulate the mathematical principles that the desired measure should satisfy. To keep it concise and sufficiently ambiguous, we will state them in terms of natural language only:

\begin{enumerate}
    \item[(i)] The complexity measure corresponds directly to the determinization difficulty of the underlying automaton.
    \item[(ii)] The complexity measure is algorithmically feasible to compute or estimate.
\end{enumerate}

Due to the famous Myhill--Nerode theorem \cite{automata-book}, the minimal deterministic automaton $\detautomaton$ is unique up to isomorphism; hence the most straightforward attempt to assess the complexity of determinization is to consider the state complexity $|\detautomaton|$. Unfortunately, this approach immediately clashes with both of our principles.
To illustrate the first issue, let us consider the following $1$-state deterministic automaton
\begin{equation}\label{eq:universal-dfa}
    \universalautomaton \coloneqq (\{\syma, \symb\}, \{ \stateq \}, \{\stateq \}, \{\stateq \}, \{(\stateq, \syma, \stateq), (\stateq, \symb, \stateq) \}),
\end{equation}
which is minimal and accepts the \defn{universal} language $\languagefsa{\universalautomaton} = \kleene{\{\syma, \symb\}}$, i.e., all words over the binary alphabet $\alphabet = \{ \syma, \symb \}$. 
\begin{theorem}[(NFA Universality \cite{pspace-survey})]\label{thm:universality}
    Given an $n$-state NFA $\automaton$, deciding whether \textup{$\languagefsa{\automaton} = \languagefsa{\universalautomaton}$} is \textup{$\pspace$}-complete.
\end{theorem}

\cref{thm:universality} shows that determinization becomes intractable even when the state complexity of the underlying NFA is exactly $1$, since any determinization procedure can be modified to decide universality. In particular, this implies that there does not exist a polynomial-time fixed-parameter tractable (FPT) determinization algorithm parametrized by state complexity $|\detautomaton|$, and hence the complexity of NFA determinization is not related to the underlying state complexity. 

We continue our journey by examining how (ii) relates to the state complexity. First, we define what a feasible estimate or approximation means in our context.
Let $\sF$ be any metric, which associates a natural number with every possible FA (for example $\sF$ could count the states, i.e., $\sF(\automaton) = |\automaton|$).

\begin{definition}\label{def:f-approx}
    Let $\automaton$ be an arbitrary NFA and let $f \colon \N \rightarrow \N$ be some function satisfying $f(i) \geq i$ for all $i \in \N$. We say that $\fapprox{m}$ is an $f$-\emph{approximation} of metric $\sF$ on $\automaton$ if 
    \begin{equation}
     \sF(\automaton) \leq \fapprox{m} \leq f(\sF(\automaton) + |\automaton|),      
    \end{equation}
    i.e., $\fapprox{m}$ is an upper bound with slackness specified by $f$.
\end{definition}

Below we focus on the particular choice for the metric---the state complexity, i.e., $\sF(\automaton) = |\detautomaton|$. Now we are ready to show that the state complexity does not allow for any reasonable approximation scheme by proving the following result.
\begin{proposition}\label{prop:pspace}
   For any polynomial $p$, it is $\pspace$-hard to compute a $p$-approximation to the state complexity.
\end{proposition}
\begin{proof}
    To show $\pspace$-hardness, we construct a polynomial reduction from the NFA universality problem. 
    Let $\automaton$ be an arbitrary NFA and let $n = |\automaton|$.
    Further, assume that there exists an algorithm that, given any NFA, produces a $p$-approximation to its state complexity.
    We show that we could use such an algorithm to decide whether $\automaton$ accepts the universal language, which is $\pspace$-hard by \cref{thm:universality}.
    
    Without loss of generality, we assume that $\alphabet = \{\syma, \symb\}$.
    Next, we construct an automaton $\automaton'$ over the alphabet $\alphabet' = \{\syma, \symb, \# \}$ with $2(n+1)$ states using a copy of $\automaton$, a copy of $\mooreautomaton$ and two copies of $\universalautomaton$ as shown in \cref{fig:minimal-pspace}. 
    Specifically, $\automaton'$ is the union of the following two NFAs:
    \begin{enumerate}
        \item[1.] We construct the first automaton as follows.
        Let $\universalautomaton$ be the 1-state minimal DFA that accepts $\kleene{\alphabet}$, and let $\mooreautomaton$ be the $n$-state Moore automaton. 
        Then, concatenate $\universalautomaton$ to  $\mooreautomaton$ by adding a $\#$-labeled transition from the unique state in $\universalautomaton$ to the initial state of $\mooreautomaton$. 
        Then, the initial state of $\mooreautomaton$ no longer serves as an initial state, and the single state in $\universalautomaton$ no longer serves as a final state.
        \item[2.] We construct the second automaton as follows.
        Let $\universalautomaton$ be another copy of the 1-state minimal DFA that accepts $\kleene{\alphabet}$. 
         Then, concatenate $\automaton$ to  $\universalautomaton$ by adding a $\#$-labeled transition from all final states in $\automaton$ to the initial state of $\universalautomaton$. 
         The unique state in $\universalautomaton$ no longer serves as an initial state, and $\automaton$ no longer has any final states.
    \end{enumerate}
    \begin{figure}[t]
    \centering
    \scalebox{.8}{
        \begin{tikzpicture}[->,>=stealth',shorten >=0.5pt,auto,node distance=15mm,
                            semithick]
        \tikzstyle{every state}=[fill=none,draw=black,text=black,minimum size=0pt,inner sep=1pt]
        
        \node[state, initial] (p1) {$p_1$};
        \node[state, above right of = p1] (p2) {$p_2$};
        \node[state, right of = p2] (p3) {$p_3$};
        \node[state, right of = p3] (p4) {$p_4$};
        \node[state, below right of = p1] (p5) {$p_5$};
        \node[fill=white, below right =2.1cm and 0.1cm of p4] (pmain){$\automaton$};
        
        \draw (p1) edge[loop below] node{$a, b$} (p1)
              (p1) edge[above left] node{$a$} (p2)
              (p1) edge[above right] node{$b$} (p5)
              (p2) edge[above] node{$a$} (p3)
              (p3) edge[above] node{$a$} (p4)
              (p4) edge[loop below] node{$a, b$} (p4);
    
        \node[state, accepting, below right= {0.75cm} and {2cm} of p4] (pu) {$p_u$};
        \node[fill=white, below =0.5cm of pu] (pmainu) {$\universalautomaton$};
        \draw (pu) edge[loop above] node{$a, b$} (pu);
        
        \node[state, above right = 4cm and 3.8cm of p1] (q1) {$q_1$};
        \node[state, right of=q1] (q2) {$q_2$};
        \node[state, right of=q2] (q3) {$q_3$};
        \node[state, right of=q3] (q4) {$q_4$};
        \node[fill=white, right of=q4] (q?) {$\ldots$};
        \node[state, accepting, right of=q?] (qn) {$q_n$};
        \node[fill=white, below=1.1cm of qn] (qmoore) {$\mooreautomaton$};
        \draw (q1) edge[loop above] node{$b$} (q1)
              (q1) edge[above] node{$a$} (q2)
              (q2) edge[above] node{$a, b$} (q3)
              (q3) edge[above] node{$a, b$} (q4)
              (q4) edge[above] node{$a, b$} (q?)
              (q?) edge[above] node{$a, b$} (qn)
              (qn) edge[bend right=1.6cm, above] node{$a$} (q2)
              (qn) edge[bend left=1.2cm, below] node{$a$} (q1);
    
        \node[state, initial, left = 2cm of q1] (qu) {$q_u$};
        \node[fill=white, below right =0.55cm and -0.25cm  of qu] (qmooreu) {$\universalautomaton$};
        \draw (qu) edge[loop above] node{$a, b$} (qu);
    
        \draw (qu) edge[above] node{$\#$} (q1);
        \draw (p4) edge[above right] node{$\#$} (pu);
          
        \node[draw, dashed, inner xsep=6mm, inner ysep = 4mm, fit=(p1) (p2) (p3) (p4) (p5) ] {};
        \node[draw, dashed, inner xsep=4mm, inner ysep = 16mm, fit=(q1) (q2) (q3) (q4) (q?) (qn)] {};
        \node[draw, dashed, inner xsep=6mm, inner ysep = 10mm, anchor=base, fit=(qu)] {};
        \node[draw, dashed, inner xsep=2mm, inner ysep = 10mm, anchor=base, fit=(pu)] {};
        \end{tikzpicture}
    }
    \caption{Construction for automaton $\automaton'$ consisting of a union of two nondeterministic automata: the upper one is a connection of $\universalautomaton$ and $\mooreautomaton$ using symbol $\#$, and the lower one is the connection of $\automaton$ and $\universalautomaton$ using symbol $\#$. In both cases the final states of the first automaton are connected to the initial states of the second one, with the corresponding initial/final labels dropped.} 
    \label{fig:minimal-pspace}
    \end{figure}
    Clearly, the first automaton accepts the language $\kleene{\alphabet}\#\languagefsa{\mooreautomaton}$ while the second accepts the language $\languagefsa{\automaton}\#\kleene{\alphabet}$. 
    Thus, $\automaton'$ accepts $(\kleene{\alphabet}\#\languagefsa{\mooreautomaton})  \cup (\languagefsa{\automaton}\#\kleene{\alphabet})$.
    On the one hand, if $\automaton$ is universal, then $\languagefsa{\automaton} = \kleene{\alphabet}$, and the union of these languages is precisely $\kleene{\alphabet}\#\kleene{\alphabet}$.
    Consequently, the state complexity of $\automaton'$ is $3$.
    On the other hand, if $\languagefsa{\automaton}$ is not universal, then there exists a word $\strx \in \kleene{\{a, b \}}$ such that $\strx \notin \languagefsa{\automaton}$, meaning that
    \begin{equation}
        \strx\# \languagefsa{\mooreautomaton} \subseteq \languagefsa{\automaton'} \quad \text{and} \quad \bigl\{ \strx\#\stry  \mid \stry \in  \setcomplement{\languagefsa{\mooreautomaton}}\bigr\} \, \cap \,\languagefsa{\automaton'} = \emptyset.
    \end{equation}
    Consider the minimal deterministic automaton for $\automaton'$, which we denote with $\detautomatonprime$. Let $q$ be the state in $\detautomatonprime$ to which the path $\apath$ with the yield $\strx\#$ leads. 
    Now, observe that the right language of $q$ is  $\languagefsa{\mooreautomaton}$.
    Because $\languagefsa{\mooreautomaton}$ has a state complexity $2^{n}$, the state complexity of $\detautomatonprime$ is at least $2^n$.
    Thus, depending on whether $\automaton$ is universal or not, the state complexity of $\automaton'$ will be either $3$ or at least $2^n$. Hence, any $p$-approximation $\fapprox{m}$ to the state complexity of $\automaton'$ satisfies 
    \begin{equation}
         \fapprox{m} \leq p(3 + |\automaton'|) = p(2n + 5)    
    \end{equation}
    in the former case and $\fapprox{m} \geq 2^n$ in the latter; since $p(2n+5) < 2^n$ for all sufficiently large $n$, any algorithm producing a $p$-approximation for the state complexity of $\automaton'$ would necessarily decide the universality of $\automaton$.
\end{proof}

A variety of alternative analytical measures for automata complexity have been introduced over time, with most notable ones being listed in \cite{complexity-survey}. A promising approach was presented in \cite{width-NFA}, which describes a novel determinization algorithm parametrized by the width of the NFA, but unfortunately it is superexponential in the worst case. More generally, there seems to be little connection between the complexity of determinization algorithms and any of the known complexity measures for NFAs, perhaps with the exception of the semigroup complexity, which we will discuss more in \cref{sec:semigroup-approach}. Hence---following the spirit of this article---we instead propose a more algorithmic approach to tackling this problem.

\section{The Algorithmic Landscape}\label{sec:algorithmic-landscape}

It seems that our desired measure of determinization complexity should be tied to the underlying determinization algorithms. Ideally, we would want to find an optimal determinization algorithm and then link some complexity measure to its running time. However, a variety of different determinization algorithms exist and, as alluded to in the previous section, all of them necessarily have a worst-case exponential runtime. Hence, to achieve our goal, we need to start by answering the following question:

\begin{quote}
    \textit{How should one compare exponential-time determinization algorithms?}
\end{quote}

\subsection{Space Complexity}

Since worst-case time complexity cannot be used as a sensible measure, a different metric to consider is the space complexity. Although this question does not seem to have appeared in the literature, an observation made by Stefan Kiefer in \cite{pspace-minimization} provides a sketch of a general NFA determinization algorithm using only polynomial space (here we rely on the classical definition where the underlying Turing machine has separate working and output tapes, allowing for exponentially large outputs). Below we present a slightly more formal statement of this idea.

\begin{theorem}[(Polynomial Space \cite{pspace-minimization})]
    There exists a determinization algorithm that transforms any $n$-state NFA $\automaton$ into a minimal equivalent DFA \textup{$\detautomaton$} with space complexity $\bigO{n^2}$. 
\end{theorem}

\begin{proof}
    The proof of this result consists essentially of two main components: the construction of an oracle for the powerset automaton using only polynomial space and the algorithm from \cite{NL-minimization} capable of performing deterministic automata minimization in $\nlclass$. 
    
    Given an NFA $\automaton = \fsatuple$, the oracle provides access to the adjacency matrices of the powerset automaton $\powerautomaton$ by receiving a query and returning a binary response. A query consists of a tuple of powerstates $\powerstate_i, \powerstate_j \in \powerset{\states}$  and a symbol $\sym \in \alphabet$, and returns $1$ if transition $\edgenoweight{\powerstate_i}{\sym}{\powerstate_j}$ exists in $\powerautomaton$, and $0$ otherwise. To check for $\edgenoweight{\powerstate_i}{\sym}{\powerstate_j}$ the oracle simply verifies whether
    \begin{equation}
        \{\stateq' \ | \ (\stateq, \sym, \stateq') \in \trans, \ \stateq \in \powerstate_i \} \cap \powerstate_j \neq \emptyset,
    \end{equation}
    which can be clearly done in $\bigO{n}$ space. Using the oracle as input to the minimization algorithm, we are able to communicate an automaton of exponential size using only polynomial space, and since the minimization algorithm is logarithmic in the input size, we conclude that $\automaton$ can be determinized on a nondeterministic Turing machine with linear space. By applying the famous Savitch's theorem \cite{savitch}, we conclude the existence of a deterministic algorithm for NFA determinization requiring only $\bigO{n^2}$ memory.
\end{proof}

While the existence of a worst-case space efficient determinization is reassuring, the construction above will always have exponential time complexity, making it highly impractical. Of course, we could attempt to improve that, but this only confirms that the space complexity alone cannot be the desired measure for our problem. 

\subsection{Comparing Algorithms}

Our next attempt at answering the original question uses the following idea: instead of comparing the worst-case time complexities of algorithms, we compare the number of steps they use (given a canonical model of computation, i.e., RAM\footnote{Time complexities of determinization algorithms are usually discussed in the context of random access machines (RAM), and converting them to Turing machines adds an extra polynomial factor.}) individually on every input NFA $\automaton$.

\begin{definition}\label{def:algorithm-compare}
 We say that a determinization algorithm $\textsc{Alg}_1$ \emph{improves} another determinization algorithm $\textsc{Alg}_2$, i.e., $\textsc{Alg}_1 \geq \textsc{Alg}_2$, if there exists a constant $c$ such that for any FA $\automaton$, the number of steps required to compute $\detautomaton$ using $\textsc{Alg}_1$ is at most $c$ times larger than the number of steps required to compute $\detautomaton$ using $\textsc{Alg}_2$.
\end{definition}

Definition \ref{def:algorithm-compare} feels quite intuitive: an algorithm $\textsc{Alg}_1$ is an improvement over $\textsc{Alg}_2$, if and only if for any input automaton $\automaton$ it is at least as fast as $\textsc{Alg}_2$ (measured up to a constant). We can easily deduce that the subset construction is an improvement over the powerset construction (as presented in \cref{sec:background-powerset}). 

Unsurprisingly, the ordering $\geq$ is not total, as illustrated by the observation below; hence there is likely no optimal determinization algorithm. 

\begin{proposition}\label{prop:incomparable}
    The subset construction and Brzozowski's algorithm are incomparable under $\geq$, i.e., neither of them is an improvement of the other.
\end{proposition}

\begin{proof}
    We first show that the subset construction does not improve Brzozowski's algorithm. Indeed, consider Moore's automaton $\mooreautomaton$ with an explicit dead state $\stateq_\varnothing$, and mark all states as final. This does not affect \cref{alg:subsetconstruction}, which means that it still produces exponentially many states. In contrast, Brzozowski's algorithm terminates immediately, since \textsc{Reverse} produces an automaton where all states are both initial and reachable, which then gets transformed into a DFA with a single state. To show the other direction, it is sufficient to exhibit some DFA $\automaton$ whose reverse $\reverse{\automaton}$ has exponential state complexity. The existence of DFAs with this property is well-known, and later in \cref{sec:inapproximability} such a DFA is explicitly constructed (see the modified Meyer--Fischer automaton $\mmfautomaton$ in \cref{fig:inverse-mf} for more details). 
\end{proof}

\begin{remark}
    Note that the ordering $\geq$ behaves like a lattice. Indeed, for any two determinization algorithms $\textsc{Alg}_1$ and $\textsc{Alg}_2$ one can construct an algorithm $\textsc{Alg}_{1 \vee 2}$ (i.e., join) which improves both of them: simply run both algorithms in parallel on the same input, and terminate as soon as one of them terminates. In a similar manner, one can construct the meet $\textsc{Alg}_{1 \wedge 2}$ by always waiting for $\textsc{Alg}_1$ and $\textsc{Alg}_2$ to terminate.
\end{remark}

Some previous work on determinization algorithms can be naturally expressed using our notion of improvement too: \cite{five-determinization} presents five different determinization algorithms (note that the authors do not require the final DFA to be minimal, hence the algorithms ought to be coupled with Hopcroft's \textsc{Minimize}), four of which are improvements of the subset construction.

\subsection{Measuring Time Complexities}

Let us briefly analyze the more granular time complexity of our determinization algorithms. First, consider the subset construction, and in particular \cref{alg:subsetconstruction}. As noted in \cite{subsethomogenous}, the number of steps it takes on $\automaton$ lies within $\mathcal{O}(\nsymbols |\ssautomaton| {|\automaton|}^2)$. The addition of \textsc{Minimize} adds a logarithmic factor $\log{|\ssautomaton|}$ (which can be bounded by $|\automaton|$), hence the final number of steps that the subset construction uses on $\automaton$ is within
\begin{equation}\label{eq:subset-time}
    \bigO{\nsymbols \left(|\ssautomaton| {|\automaton|}^2 + |\ssautomaton|\log{|\ssautomaton|}\right)} = \bigO{\nsymbols |\ssautomaton| {|\automaton|}^2}.
\end{equation}

Brzozowski's algorithm, as formulated in \cref{sec:background-brzozowski}, effectively consists of two runs of the subset construction: in the first stage it produces the subset automaton for $\reverse{\automaton}$ and, in the second stage, it produces another subset automaton, which, by \cref{thm:brzozowski}, coincides with the minimal DFA $\detautomaton$. Denote by $\brautomaton$ the subset automaton for $\reverse{\automaton}$, then the number of steps that Brzozowski's algorithm uses on $\automaton$ lies within
\begin{equation}\label{eq:brzozowski-time}
    \bigO{\nsymbols \left({|\automaton|}^2 |\brautomaton|  + {|\brautomaton|}^2 |\detautomaton|\right)}.
\end{equation}

\begin{definition}\label{def:alg-approximation}
    A determinization algorithm \textsc{Alg} is said to admit an \emph{$f$-approximation} if the metric $\sF$ measuring the number of steps (i.e., time complexity) of \textsc{Alg} on a given input $\automaton$ admits an $f$-approximation (as in Definition \ref{def:f-approx}).
\end{definition}

As observed earlier, the metric $\sF$ from Definition \ref{def:alg-approximation} both for the subset construction and Brzozowski's algorithm is dominated by the sizes of the corresponding subset automata. In what follows, we will attempt to better understand this metric and analyze its asymptotic behavior and computational aspects.

\section{Time Complexity is Inapproximable}\label{sec:inapproximability}

In this section we demonstrate that the performance of almost all efficient determinization algorithms is extremely difficult to estimate, with the key findings presented in \cref{thm:inapproximability} and \cref{thm:inapp-brzozowski} below. Before stating these results, we begin by making two straightforward observations about \textsc{SubsetConstr}.

\begin{lemma}\label{lem:all-states}
    The NFA $\automaton = \fsatuple$ is universal if and only if every state $\powerstate$ in the subset automaton \textup{$\ssautomaton$} (produced by \cref{alg:subsetconstruction}) contains a final state from $\automaton$.
\end{lemma}
\begin{proof}
    To show the left implication, observe that if all states generated by applying the subset construction contain a final state, then all states in $\ssautomaton$ are accepting by design of \cref{alg:subsetconstruction}; hence, $\ssautomaton$ accepts a universal language. 
    To show the right implication, observe that if there does exist a state $\powerstate$ generated by \Cref{alg:subsetconstruction} that does \emph{not} contain a final state, then, since all states in $\ssautomaton$ are accessible, there exists a path $\apath$ with yield $\stry$ from $\powerstate_\initial$ to $\powerstate$, which means that $\stry \notin \languagefsa{\ssautomaton}$. 
\end{proof}

\begin{lemma}\label{lemma:MF-fixpoint}
    Let \textup{$\mfautomaton = (\alphabet, \mfstates, \initial, \final, \trans_\textsc{mf})$} be an $n$-state Meyer--Fischer automaton.
    Consider the powerstate \textup{$\powerstate_1 = \mfstates$} in its powerset automaton \textup{$\powerautomaton$}.
    Then, all transitions leaving $\powerstate_1$ in \textup{$\powerautomaton$} are self-loops. 
\end{lemma}
\begin{proof}
    Because being in the state $\powerstate_1$ corresponds to simultaneously being in all the states $\{\stateq_1, \ldots, \stateq_n\}$ of $\mfautomaton$,\footnote{See \cref{fig:Moore-NFA}.} the claim follows by observing that every state in $\mfautomaton$ has at least one incoming transition with label $\syma$ and one with $\symb$.
    Thus, once one lands in $\powerstate_1$, one will always transition back to all states in $\powerstate_1$. 
\end{proof}

Equipped with these lemmas, we are now ready to discuss the main results of this section.

\begin{theorem}\label{thm:inapproximability}
    For any polynomial $p$, it is $\pspace$-hard to compute a $p$-approximation for the subset construction or any  improvement thereof.
\end{theorem}

\begin{proof}
    We begin by showing the claim only for the subset construction. 
    To show $\pspace$-hardness of computing a $p$-approximation, we again construct a polynomial reduction from the NFA universality problem.
    Assume that we have access to an algorithm which, given any $\automaton = \fsatuple$, computes a $p$-approximation $\widehat{m}$ to the size of $\automaton$'s subset automaton.
    Let $n = |\automaton |$ and let $d$ be the degree of the polynomial $p$ and without loss of generality assume that $\alphabet = \{\syma, \symb\}$. 
    Given a fixed positive integer $k \in \N$ consider an $n(k+1)$-state Meyer--Fischer automaton $\mfautomaton$ over $\alphabet$.
    As with the $n(k+1)$-state Moore automaton, the $n(k+1)$-state Meyer--Fischer $\mfautomaton$ has a state complexity of $2^{n(k+1)}$.
    In addition, the Meyer--Fischer automaton has a unique initial state $p_1$ which is also the unique accepting state, as shown in the top part of \cref{fig:subset-pspace}.
    
    Without loss of generality assume that $\automaton$ has an explicit dead state $\stateq_{\varnothing}$, i.e., if some state $\stateq \in Q$ is missing an explicit transition for some symbol $\sym \in \Sigma$, we augment $\automaton$ with transition $\edgenoweight{\stateq}{\sym}{\stateq_{\varnothing}}$. 
    In other words, $\automaton$'s transition function is total.
    \begin{figure}[t]
    \centering
    \scalebox{.8}{
    \begin{tikzpicture}[->,>=stealth',shorten >=0.5pt,auto,node distance=15mm,
                        semithick]
    \tikzstyle{every state}=[fill=none,draw=black,text=black,minimum size=0pt,inner sep=1pt]
    
    \node[state, initial] (p1) {$q_1$};
    \node[state, above right of = p1] (p2) {$q_2$};
    \node[state, right of = p2] (p3) {$q_3$};
    \node[state, very thick, right of = p3] (p4) {$q_4$};
    \node[state, below right of = p1] (p5) {$q_5$};
    \node[state, right of = p5] (p6) {$q_{\text{\scalebox{0.75}{$\varnothing$}}}$};
    \node[fill=white, below right=2.1cm and 0.2cm of p4]{$\automaton$};
    
    \draw (p1) edge[loop below] node{$a, b$} (p1)
          (p1) edge[above left] node{$a$} (p2)
          (p1) edge[above right] node{$b$} (p5)
          (p2) edge[above] node{$a$} (p3)
          (p3) edge[above] node{$a$} (p4)
          (p4) edge[loop below] node{$a, b$} (p4)
          (p2) edge[right] node{$b$} (p6)
          (p3) edge[right] node{$b$} (p6)
          (p5) edge[above] node{$a, b$} (p6)
          (p6) edge[loop right] node{$a, b$} (p6);
    
    \node[state, accepting, above right = 3.5cm and 3.8cm of p1] (q1) {$p_1$};
    \node[state, above right = {12mm} and {3mm} of q1] (q2) {$p_2$};
    \node[state, above right of=q2] (q3) {$p_3$};
    \node[state, right of=q3] (q4) {$p_4$};
    \node[fill=white, below right of=q4] (q?) {$\ldots$};
    \node[state, right = {43mm} of q1] (qn) {$p_t$};
    \node[fill=white, below right=0.6cm and 0.1cm of qn]{$\mfautomaton$};
    \draw (q1) edge[in=85,out=115,loop, above] node{$b$} (q1)
          (q2) edge[loop above] node{$b$} (q2)
          (q3) edge[loop above] node{$b$} (q3)
          (q4) edge[loop above] node{$b$} (q4)
          (qn) edge[loop right] node{$b$} (qn)
          (q1) edge[above left] node{$a$} (q2)
          (q2) edge[above left] node{$a$} (q3)
          (q3) edge[above] node{$a$} (q4)
          (q4) edge[above right] node{$a$} (q?)
          (q?) edge[above right] node{$a$} (qn)
          (qn) edge[above] node{$a, b$} (q1)
          (q2) edge[bend left=1cm,left] node{$b$} (q1)
          (q3) edge[bend left=1cm,left] node{$b$} (q1)
          (q4) edge[bend left=1cm,above] node{$b$} (q1)
          ;
    
    \draw (p1) edge[left, bend left=1.3cm] node{$\#$} (q1)
          (p2) edge[left, bend left=1cm] node{$\#$} (q1)
          (p3) edge[left, bend left=0.4cm] node{$\#$} (q1)
          (p5) edge[left, bend left=0.7cm] node{$\#$} (q1)
          (p6) edge[left, bend left=0.1cm] node{$\#$} (q1);
    
    \draw (p4) edge[above right, bend right=0.1cm] node{$\#$} (q1)
          (p4) edge[right, bend right=0.4cm] node{$\#$} (q2)
          (p4) edge[right, bend right=0.7cm] node{$\#$} (q3)
          (p4) edge[right, bend right=0.9cm] node{$\#$} (q4)
          (p4) edge[below right, bend right=0.7cm] node{$\#$} (qn);
              
    \node[draw, dashed, inner xsep=6mm, inner ysep = 4mm, fit=(p1) (p2) (p3) (p4) (p5) (p6)] {};
    \node[draw, dashed,inner xsep=8mm, inner ysep = 10mm, fit=(q1) (q2) (q3) (q4) (q?) (qn)] {};
    \end{tikzpicture}
    }
    \caption{The construction of the automaton $\automaton'$ from an arbitrary NFA $\automaton$ with an explicit dead state $q_\varnothing$ and the $n(k+1)$-state Meyer--Fischer automaton $\mfautomaton$.
    All non-final states of $\automaton$ are connected to the initial state $p_1$ of $\mfautomaton$ using symbol $\#$ and all final states ($q_4$ is the only one in $\automaton$) are connected to all states of $\mfautomaton$ using the same symbol $\#$.
    In the diagram, $t = n(k+1)$.
    }
    \label{fig:subset-pspace}
    \end{figure}
    
    Next, we create an automaton $\automaton'$ over the alphabet $\alphabet' = \{\syma, \symb, \#\}$ by conjoining $\automaton$ and $\mfautomaton$ as follows (see \cref{fig:subset-pspace}):
    \begin{enumerate}
        \item[1.] We add $\#$-labeled transitions from all nonfinal states of $\automaton$ to $\mfautomaton$'s initial state $p_1$.
        The state $p_1$ is no longer initial in $\automaton'$.
        \item[2.] We add $\#$-labeled transitions from all final states in $\automaton$ to all states of $\mfautomaton$. The final states from $\final$ are no longer final in $\automaton'$. 
    \end{enumerate}
    Next, we analyze an application of the subset construction to $\automaton'$ by breaking its computation into two stages.
    In the first stage, we only traverse transitions with symbols $\syma$ and $\symb$.
    This corresponds \emph{exactly} to running \cref{alg:subsetconstruction} on $\automaton$.
    After the first stage is completed, i.e., after no new states can be generated, we allow \cref{alg:subsetconstruction} to consider other symbols from $\alphabet'$. 
    Note that our decomposition into two stages is little more than an enforcement of a specific ordering of the computation performed by \cref{alg:subsetconstruction} when determinizing $\automaton'$---it does not change \emph{what} is computed, since the set of powerstates generated by \cref{alg:subsetconstruction} does not depend on the order in which they are explored.
    We denote the resulting subset automaton as $\ssautomatonprime$.

    Let us observe the (incomplete) subset automaton after the first stage. 
    We first consider the case where $\automaton$ is universal.
    Due to Lemma \ref{lem:all-states}, all states in the (incomplete) subset automaton contain at least one final state from $\automaton$. 
    Hence, in the second stage of the subset construction, any $\#$-labeled transition leads to the powerstate $\powerstate_1 = \{p_1, \ldots, p_t\}$, where, due to Lemma \ref{lemma:MF-fixpoint}, we either remain forever (for $\alphabet$-labeled transitions) or transition into the powerstate $\emptyset$ (for $\#$-labeled transitions).
    This means that exactly 2 additional states are produced in the second stage.  
    
    Next, we consider the case where $\automaton$ is \emph{not} universal and let $\strx \in \kleene{\{a, b \}} \setminus \languagefsa{\automaton}$ be an arbitrary word not in the language. Observe that in this case
    \begin{equation}\label{eq:mf-is-large}
        \strx\# \languagefsa{\mfautomaton} \subseteq \languagefsa{\automaton'} \quad \text{and} \quad \bigl\{ \strx\#\stry  \mid \stry \in  \setcomplement{\languagefsa{\mfautomaton}}\bigr\} \, \cap \,\languagefsa{\automaton'} = \emptyset.
    \end{equation}
    since every path with yield $\strx$ ends in a non-final state of $\automaton$, meaning that $\strx \#$ leads exclusively to $p_1$ in $\automaton'$.
    Now consider the minimal deterministic automaton for $\automaton'$, which we denote with $\detautomatonprime$ and let $\stateq'$ be the state in $\detautomatonprime$ to which the path $\apath$ with the yield $\strx\#$ leads. 
    Observe that the right language of $\stateq'$ is $\languagefsa{\mfautomaton}$, meaning that the number of states in $\detautomatonprime$ satisfies $|\detautomatonprime| \geq 2^{n(k+1)}$.
    
    In summary, if $\automaton$ is universal, then the subset automaton $\ssautomatonprime$ will have exactly $2$ more states than $\ssautomaton$, while if $\automaton$ is not universal, it will require at least $2^{n(k+1)}$ states. 
    Moreover, since $\automaton$ has only $n$ states, the size of $\ssautomaton$ will be at most $2^{n}$. Thus, if $\automaton$ is universal, we have $|\ssautomatonprime| \leq 2^n + 2$. 
    And, if $\automaton$ is not universal, we have $|\ssautomatonprime| \geq |\detautomatonprime| \geq 2^{n(k+1)}$. 
    It follows that if $\automaton$ is universal, any $p$-approximation $\widehat{m}$ for $m = |\ssautomatonprime|$ will satisfy
    \begin{equation}
        \widehat{m} \leq p(m + |\automaton'|) \in \bigO{{(m+n(k+2))}^d} = \bigO{{(2^{n} + 2 + n(k+2))}^d}.
    \end{equation}
    And, if $\automaton$ is not universal, $\widehat{m} \geq 2^{n(k+1)}$.
    We choose $k \gg d$, the degree of $p$. If we assume that $n$ is sufficiently large then, by comparing $\widehat{m}$ and $2^{n(k+1)}$, we can decide the universality of $\automaton$. 
    
    Due to \cref{eq:subset-time}, the number of steps in the subset construction is proportional to $m$ by a factor of $|\automaton'|^2 = {n^2(k+2)^2}$, hence a polynomial approximation for the number of steps can be converted into a polynomial approximation for $m = |\ssautomatonprime|$.
    
    We conclude that it is \pspace-hard to compute a $p$-approximation for the subset construction on general NFAs. This argument also holds for any improvements of the subset construction, since the number of steps can be smaller than $|\ssautomatonprime|$ in case $\automaton$ is universal, but has to be at least $|\detautomatonprime| \geq 2^{n(k+1)}$ in case $\automaton$ is not. Thus, any $p$-approximation decides universality of $\automaton$ for \textit{any} improvement of the subset construction. 
\end{proof}

\begin{theorem}\label{thm:inapp-brzozowski}
    For any polynomial $p$, it is $\pspace$-hard to compute a $p$-approximation for Brzozowski's algorithm or any improvement thereof.
\end{theorem}

\begin{proof}
    Let $\automaton = \fsatuple$ be an FA with $n = |\automaton|$, $p$ an arbitrary polynomial of degree $d$, and let $k \in \N$ be some positive integer. We will construct $\automaton'$ similarly to \cref{thm:inapproximability}, but instead of adjoining the $n(k+1)$-state Meyer--Fischer automaton $\mfautomaton$ we will adjoin the modified Meyer--Fischer automaton, denoted with $\mmfautomaton$, as presented in \cref{fig:inverse-mf}. We designed $\mmfautomaton$ in such a way that $\languagefsa{\mfautomaton} = \languagefsa{\mmfautomaton}$ but, in contrast to the original, $\mmfautomaton$ is co-deterministic (its reverse is given explicitly in \cref{fig:inverse-mf}). 
    
    Notice that Lemma \ref{lemma:MF-fixpoint} also holds for $\mmfautomaton$, thus, analogously to \cref{thm:inapproximability}, if $\automaton$ \emph{is} universal, no matter when we take a transition with label $\#$, we will stay forever in $\powerstate_1 = \{p_1, \ldots, p_t, p_b \}$, which is accepting. Thus, the underlying language satisfies
    \begin{equation}
         \languagefsa{\detautomatonprime} = \languagefsa{\automaton'} = \languagefsa{\automaton} \# \kleene{\{\syma, \symb \}} = \kleene{\{\syma, \symb\}} \# \kleene{\{\syma, \symb\}}.
    \end{equation}
    Similarly, if $\automaton$ is \emph{not} universal, then the state complexity $|\detautomatonprime|$ is at least $2^{n(k+1)}$ due to \cref{eq:mf-is-large}.
    
     Our goal is to show that for any universal $\automaton$, Brzozowski's algorithm will not produce too many states on $\automaton'$. To achieve that we will analyze its execution using two stages: first reversing $\automaton'$, running \cref{alg:subsetconstruction} and reversing the result and, second, running \cref{alg:subsetconstruction} again.
    \begin{nclaim1}
        The number of states after the first stage of Brzozowski's algorithm on $\automaton'$ is at most $k2^{n+2}$.
    \end{nclaim1}
    The first stage of Brzozowski's algorithm corresponds to running \cref{alg:subsetconstruction} on $\reverse{(\automaton')}$ which is, effectively, the concatenation of $\reverse{(\mmfautomaton)}$ with $\reverse{\automaton}$ using $\#$ (recall that the copy of $\automaton$ within $\automaton'$ has no final states). 
    
    First, \cref{alg:subsetconstruction} is launched on initial states from $\reverse{(\mmfautomaton)}$, but, while exploring only $\alphabet$-labeled transitions, no new states will be created because $\reverse{(\mmfautomaton)}$ is already deterministic. As soon as $\#$-labeled transitions are encountered, the situation changes: all $\#$-labeled transitions from $\reverse{(\mmfautomaton)}$ lead either into the powerstate $\powerstate = \states$ or $\powerstate' = \final$ in $\reverse{\automaton}$. In the spirit of \cref{thm:inapproximability} we enforce the following ordering on the execution of \cref{alg:subsetconstruction}:
    \begin{figure}[t]
    \begin{center}
        \begin{tikzpicture}[->,>=stealth',shorten >=0.5pt,auto,node distance=15mm, scale=0.8, transform shape]
        \tikzstyle{every state}=[fill=none,draw=black,text=black,minimum size=0pt,inner sep=2.5pt]
            \node[state, accepting, initial] (q1) {$p_1$};
            \node[state, above right = {12mm} and {3mm} of q1] (q2) {$p_2$};
            \node[state, above right of=q2] (q3) {$p_3$};
            \node[state, left of=q2, initial] (qf) {$p_{b}$};
            \node[state, right of=q3] (q4) {$p_4$};
            \node[fill=white, below right of=q4] (q?) {$\ldots$};
            \node[state, right = {43mm} of q1] (qn) {$p_t$};

            \draw (q2) edge[loop above] node{$b$} (q2)
                  (q3) edge[loop above] node{$b$} (q3)
                  (q4) edge[loop above] node{$b$} (q4)
                  (qf) edge[loop above] node{$a, b$} (qf)
                  (qn) edge[in=50,out=80,loop, above] node{$b$} (qn)
                  (q1) edge[above left] node{$a$} (q2)
                  (q2) edge[above left] node{$a$} (q3)
                  (q3) edge[above] node{$a$} (q4)
                  (q4) edge[above right] node{$a$} (q?)
                  (q?) edge[above right] node{$a$} (qn)
                  (qn) edge[above] node{$a$} (q1)
                  (qf) edge[left] node{$b$} (q1);
        \end{tikzpicture}
        \hfill
        \begin{tikzpicture}[->,>=stealth',shorten >=0.5pt,auto,node distance=15mm, scale=0.8, transform shape]
        \tikzstyle{every state}=[fill=none,draw=black,text=black,minimum size=0pt,inner sep=2.5pt]
            \node[state, accepting, initial] (q1) {$p_1$};
            \node[state, above right = {12mm} and {3mm} of q1] (q2) {$p_2$};
            \node[state, above right of=q2] (q3) {$p_3$};
            \node[state, left of=q2, accepting] (qf) {$p_{b}$};
            \node[state, right of=q3] (q4) {$p_4$};
            \node[fill=white, below right of=q4] (q?) {$\ldots$};
            \node[state, right = {43mm} of q1] (qn) {$p_t$};

            \draw (q2) edge[loop above] node{$b$} (q2)
                  (q3) edge[loop above] node{$b$} (q3)
                  (q4) edge[loop above] node{$b$} (q4)
                  (qf) edge[loop above] node{$a, b$} (qf)
                  (qn) edge[in=50,out=80,loop, above] node{$b$} (qn)
                  (q2) edge[above left] node{$a$} (q1)
                  (q3) edge[above left] node{$a$} (q2)
                  (q4) edge[above] node{$a$} (q3)
                  (q?) edge[above right] node{$a$} (q4)
                  (qn) edge[above right] node{$a$} (q?)
                  (q1) edge[above] node{$a$} (qn)
                  (q1) edge[left] node{$b$} (qf);
        \end{tikzpicture}
        \end{center}
        \caption{Modified Meyer--Fischer automaton $\mmfautomaton$ with an additional state $p_b$ (left) satisfying $\languagefsa{\mmfautomaton} = \languagefsa{\mfautomaton}$ for $\mfautomaton$ with $t=n(k+1)$. Its reverse, $\reverse{(\mmfautomaton)}$ (right), is deterministic.}
        \label{fig:inverse-mf}
    \end{figure}
    \begin{enumerate}
        \item[1.] Take any $\#$-transition from $\reverse{(\mmfautomaton)}$ into $\powerstate$ and find all reachable powerstates 
        \item[2.] Take any $\#$-transition from $\reverse{(\mmfautomaton)}$ into $\powerstate'$ and find all reachable powerstates 
    \end{enumerate}
    Notice that after both 1. and 2. are completed, there are no new states for \cref{alg:subsetconstruction} to discover within $\reverse{(\automaton')}$, hence it has to terminate. Furthermore, the number of states discovered in 1. is at most $2^n$, as is the number of states discovered in 2.---indeed, both  1. and 2. can be interpreted as running two instances of \cref{alg:subsetconstruction} on two separate copies of $n$-state NFA $\reverse{\automaton}$. Hence, after the first stage on $\automaton'$ we obtain some automaton $\brautomatonprime$ satisfying
    \begin{equation}
        |\brautomatonprime| \, \leq \, |\mmfautomaton| + 2^n + 2^n \, = \, 2^{n+1} + n(k+1) + 1 \, \leq \, k2^{n+2},
    \end{equation}
    which concludes the claim.
    
    In the second stage, Brzozowski's algorithm executes the subset construction on $\brautomatonprime$, but this time the subset automaton for $\brautomatonprime$ is the minimal DFA $\detautomatonprime$ with exactly $3$ states. Thus, if $\automaton$ \emph{is} universal, then, due to \cref{eq:brzozowski-time}, the total number of steps will be proportional to at most
    \begin{equation}
        {|\automaton'|}^2|\brautomatonprime| + |\detautomatonprime|{|\brautomatonprime|}^2 \, \leq \, {(n(k+2) + 1)}^2 k 2^{n+2} + 3 k^2 2^{2n + 4} \, \leq \, n^2 k^4 2^{9n}.
    \end{equation}
    On the other hand, if $\automaton$ is \emph{not} universal, any algorithm will require at least $|\detautomatonprime| \geq 2^{n(k+1)}$ many steps. Following the argument from \cref{thm:inapproximability}, by choosing $k \gg d$, any $p$-approximation for Brzozowski's algorithm (or any improvement thereof) on $\automaton'$ can be used to decide universality of $\automaton$.
\end{proof}

As a bonus, we show a somewhat unexpected consequence---the performance of the subset construction and Brzozowski's algorithm remains intractable even when the exact state complexity is known, and, as a consequence, it is also intractable to predict how much overhead these determinization algorithms require in addition to producing the minimal DFA.

\begin{corollary}\label{corr:known-state}
   For any polynomial $p$, it remains $\pspace$-hard to compute a $p$-approximation for the subset construction or Brzozowski's algorithm even when the algorithm is applied to input NFAs whose state complexity is known in advance (in fact, equal to $1$).
\end{corollary}

\begin{proof}
    To show the claim, we modify the construction from \cref{thm:inapproximability}. Let $\automaton^\cup$ denote the union automaton of $\automaton'$ and $\universalautomaton$ from \cref{eq:universal-dfa}, made universal over the alphabet $\alphabet \cup \{ \# \}$, and denote the unique state of $\universalautomaton$ by $\stateq_u$. Since \textup{$\languagefsa{\automaton^\cup} = \kleene{(\alphabet \cup \{\#\})}$}, the state complexity of $\automaton^\cup$ is known and is $1$; the same holds for its reverse \textup{$\reverse{(\automaton^\cup)}$}. It is easy to see that the execution path of both subset construction is equivalent between $\automaton'$ and $\automaton^\cup$, and the same holds for $\reverse{(\automaton')}$ and $\reverse{(\automaton^\cup)}$ in case of Brzozowski's algorithm.
\end{proof}

\begin{corollary}
    Any $p$-approximation for the overhead of the subset construction on a given NFA $\automaton$, i.e., the difference \textup{$|\ssautomaton| - |\detautomaton|$}, is $\pspace$-hard to compute.
\end{corollary}

We informally summarize the findings in this section as follows: no metric can accurately measure the performance of any good determinization algorithm, sometimes even in the presence of additional advice. In the remaining part of this paper we therefore drop the accuracy requirement and instead look for metrics that can provide \emph{any} information about potential performance of determinization.

\section{Semigroup Complexity}\label{sec:semigroup-approach}

Due to \cref{thm:inapproximability} and \cref{thm:inapp-brzozowski}, no reasonable approximation for the complexity of determinization algorithms is possible (assuming $\pspace \neq \pclass$) or, in other words, no potential metric can be both
\begin{enumerate}
    \item[(i)] a reasonably accurate approximation
    \item[(ii)] computable in polynomial time for all NFAs.
\end{enumerate}

We will take a middle path between (i) and (ii) and, in what follows, focus on bounding the size of the subset automaton $\ssautomaton$, since it plays a key role in many determinization algorithms and, more generally, can serve as a simple practical indicator. One way of achieving such a bound is by looking at \defn{semigroup complexity} \cite{complexity-survey}, a metric measuring the size of the transition monoid $\transmonoid{\automaton}$.

\begin{theorem}[(Semigroup Complexity)]\label{thm:semigroup}
    Given an NFA $\automaton$, its subset automaton \textup{$\ssautomaton$} satisfies \textup{$|\ssautomaton| \leq |\transmonoid{\automaton}|$}.
\end{theorem}

\begin{proof}
    Following \cref{alg:subsetconstruction} we conclude that since every element of $\ssstates$ is the image of at least one relation in $\transmonoid{\automaton}$, we have $|\ssstates| \leq | \transmonoid{\automaton}|$.
\end{proof}

We remark that in general transition monoids tend to be very large, sometimes much larger than the actual number of states in the FA. For $n$-state deterministic FAs over a binary alphabet, the syntactic monoid recognizing the same language can have size up to $n^n \left(1 - 2 / \sqrt{n} \right)$ (see \cite{holzer2004deterministic} for a deeper discussion). Furthermore, there is no general way of computing the exact semigroup complexity, hence, in what follows, we will restrict ourselves to subclasses of NFAs for which this is feasible.

\subsection{Unary Automata}\label{subsec:unary}

For any unary NFA $\automaton_1$ let $A$ be its unique transition matrix. Then the bound on $|\transmonoid{\automaton_1}|$ relies solely on the \defn{precedence graph} $\graph(A)$, i.e., a directed graph with $|\automaton|$ vertices and edges defined by the transition matrix $A$.

\begin{definition}\label{def:cyclicity}
    Given a directed graph $\mathcal{\graph}$ the \defn{cyclicity} of $\graph$, denoted with $c(\graph)$, is the least common multiple of cyclicities of all maximal strongly connected components of $\graph$, where the cyclicity of a strongly connected component $\sC$ is the greatest common divisor of the lengths of all cycles in $\sC$.
\end{definition}

The strongly connected components in any directed graph can be identified in linear time \cite{tarjan-scc}, therefore we can get a trivial upper bound for the cyclicity as the product of the cardinalities of all strongly connected components, since the cyclicity of each strongly connected component is at most its cardinality. The following result from semigroup theory allows us to directly bound the semigroup complexity.

\begin{proposition}[(Index and Period \cite{Markowsky_1976})]\label{prop:monoid-one-letter}
    Let $\automaton_1 = (\{ \sym \}, \states_1, \initial_1, \final_1, \trans_1)$ be an $n$-state unary automaton with transition matrix $A$. Then 
    \begin{equation}
        c(\graph(A)) \leq |\transmonoid{\automaton_1}|  \leq  c(\graph(A)) + n^2 - 2n + 2.
    \end{equation}
\end{proposition}

\subsection{Permutation Automata}\label{subsec:permutation}

An NFA $\automaton = \fsatuple$ is a \defn{permutation automaton} if for every letter $\sym \in \alphabet$ the transition relation $\trans$ defines a permutation of the set of states $\states$. Note that in this case $\transmonoid{\automaton}$ is a permutation group generated by the corresponding transition matrices. Due to the famous Schreier--Sims algorithm \cite{schreier-sims} from computational group theory, the size of any permutation group can be computed in quasi-quadratic time.

\subsection{Commutative Automata}\label{subsec:commutative}

An automaton $\automaton$ is \emph{commutative} if all transition matrices in $\{ T^{(\sym)} \ | \ \sym \in \alphabet \}$ commute with each other. Its underlying language $\languagefsa{\automaton}$ is closed under letter permutations, i.e., for any word $\str \in \languagefsa{\automaton}$ and any permutation $\pi$, we have $\pi(\str) \in \languagefsa{\automaton}$. The following result follows immediately.

\begin{proposition}\label{prop:commutative}
    Given a commutative automaton $\automaton$, define for any $\sym \in \alphabet$ the transition monoid generated by $T^{(\sym)}$ as $\transmonoid{T^{(\sym)}}$. Then:
    \begin{equation}
        |\transmonoid{\automaton}| \leq \prod_{\sym \in \alphabet} |\transmonoid{T^{(\sym)}}|
    \end{equation}
\end{proposition}

In particular, this implies that any bound from \cref{subsec:unary} can be applied individually to each unary subautomaton of $\automaton$.

\subsection{Dense Automata}\label{subsec:dense-automata}

Another interesting class of NFAs to consider is NFAs with many nondeterministic transitions. From the case of Moore's automaton (see \cref{fig:Moore-NFA}) we know that adding a single nondeterministic transition---i.e., a transition whose addition makes the automaton no longer deterministic---can have an exponential effect on the state complexity of the NFA. In this section we show that, in some sense, the opposite is true, i.e., that in the presence of many transitions (i.e., when $|\trans|$ is large compared to $\nstates$), the semigroup complexity (and hence the determinization) will likely stay sub-exponential.

\begin{definition}\label{def:indecomposable}
    An $n \times n$ Boolean matrix $B$ is \emph{$r$-indecomposable} if and only if for any non-zero Boolean (row) vector $v$ with $|v|$ non-zero entries it holds:
    \begin{equation}
        |vB| \geq \min{\{n, \ | v | + r \}}
    \end{equation}
\end{definition}

This property has a direct relation to semigroup complexity: indeed, it is easy to see that any product of $n/r$ arbitrary $n \times n$ $r$-indecomposable Boolean matrices results in a \textbf{1}-block matrix. We will use the notation $1-o(1)$ to denote the probability which becomes arbitrarily close to $1$ as the parameter $n \to \infty$. The subsequent result highlights the key fact linking indecomposable matrices to NFAs with many transitions.

\begin{lemma}[(Dense Matrices \cite{kim1978two})]\label{lem:dense-matrix}
    An $n \times n$ Boolean matrix picked uniformly at random from the set of matrices with exactly $(1 + \epsilon + r)n \log{n}$ non-zero entries will be $r$-indecomposable with probability $1 - o(1)$ for an arbitrary $\epsilon > 0$.
\end{lemma}

By combining the properties of indecomposable matrices bounding the size of the transition monoid with the distribution result on dense matrices from Lemma \ref{lem:dense-matrix}, we can formulate the final statement.

\begin{proposition}
    Let $d \in \N_+$ be a constant, $\alphabet$ some fixed alphabet and $\automaton$ an NFA over $\alphabet$ sampled uniformly at random from the set of all NFAs where for every $\sym \in \alphabet$ the transition matrix $T^{(\sym)}$ contains at least $n^2 / d$ transitions. Then, with probability $1 - o(1)$ it holds:
    \begin{equation}
        |\transmonoid{\automaton}| \leq n^{d \log{\nsymbols} + o(1)}
    \end{equation}
\end{proposition}

\begin{proof}
    Assume that all transition matrices in $\automaton$ are $r$-indecomposable for some $r \in \N$, and consider a product of arbitrary $\left\lceil \frac{n-1}{r} \right\rceil$ matrices from $\{ T^{(\sym)} \ | \ \sym \in \alphabet \}$. Due to Definition \ref{def:indecomposable}, each row of the resulting product will have at least $\min{\{n, \ 1 + \left\lceil \frac{n-1}{r} \right\rceil\cdot r\}} = n$ entries, thus we will get a \textbf{1}-block matrix. Thus, all elements of $\transmonoid{\automaton}$ with at least one zero entry will be the products of at most $\left\lceil \frac{n - 1}{r} \right\rceil - 1$ transition matrices. This means that the cardinality of $\transmonoid{\automaton}$ is bounded by the number of words over the alphabet $\alphabet$ with length of at most $\left\lceil \frac{n - 1}{r} \right\rceil - 1$, i.e.
    \begin{equation}
        |\transmonoid{\automaton}| - 1 \leq \!\!\! \!\!\!\sum_{i = 0}^{\lceil (n-1)/r \rceil - 1} \!\!\!\nsymbols^i \leq \frac{\nsymbols^{\lceil (n-1) / r \rceil}}{\nsymbols - 1} \leq {2{\nsymbols}^{(n-1) /r}}
    \end{equation}
    By Lemma \ref{lem:dense-matrix}, with probability $1 - o(1)$ all randomly sampled transition matrices of $\automaton$ will be $r$-indecomposable with $r = \frac{n}{d \log{n}} - (1 + \epsilon)$ (note that our matrices have at least $n^2/d$ non-zero entries, hence we are sampling with $\epsilon$ of at least $n/\log{n}$). Thus, by performing the substitution we get
    \begin{equation}
        |\transmonoid{\automaton}| \leq 2{\nsymbols}^\frac{d(n-1)\log{n}}{n - (1 + \epsilon)d\log{n}} \leq \nsymbols^{d \log{n} + o(1)}
    \end{equation}
    with probability $1 - o(1)$.
\end{proof}

We could not find any other major classes of automata where meaningful inference on their semigroup complexity can be performed. In the subsequent sections we attempt to mitigate this by modifying our proposed complexity metric.

\section{Subset Complexity}\label{sec:subset-complexity}

Another natural way to bound the size of the subset automaton is to consider individual ranges $\range{T^{(\sym)}}$ for Boolean transition matrices in $\{ T^{(\sym)} \ | \ \sym \in \alphabet \}$. Since every state in the subset automaton corresponds to a characteristic vector over $\states$, it has to be in the image of some transition matrix $T^{(\sym)}$ for $\sym \in \alphabet$. Hence, the size of their combined images is an upper bound on the size of the subset automaton.
\begin{lemma}[(Range Complexity)]\label{prop:range}
   Let $\automaton$ be an NFA with a set of transition matrices $\{ T^{(\sym)} \ | \ \sym \in \alphabet \}$. 
   Then its subset automaton \textup{$\ssautomaton$} satisfies
    \textup{
    \begin{equation}
        |\ssautomaton| \leq 1 +\sum_{\sym \in \Sigma} \left|\range{T^{(\sym)}} \right|,
    \end{equation}
    }
     where $\range{T}$ is the range of the matrix $T$ over the Boolean semifield $\B$. 
\end{lemma}

\begin{proof}
    Let $\ssautomaton = (\alphabet, \ssstates, \initial, \ssfinal, \sstrans)$. 
    Consider any $\powerstate_J \in \ssstates$
    and let $\sym_1 \cdots \sym_N \in \kleene{\alphabet}$ be a word such that $\powerstate_J = \trans(\powerstate_I, \sym_1 \cdots \sym_N)$.
    Then, $\powerstate_J$'s characteristic vector is given by
    \begin{equation}
        J = I^\top T^{(\sym_1)} T^{(\sym_2)} \cdots T^{(\sym_N)}.
    \end{equation}
    Moreover, because this holds for any $J$, we have
    \begin{equation}
        \ssstates \setminus \{ \powerstate_\initial \} \subseteq  \bigcup_{\sym \in \alphabet} \range{T^{(\sym)\top}}.
    \end{equation}
    Note that the row space and the column space of a Boolean matrix have the same cardinality, i.e., $|\range{T^{(\sym)\top}}| = |\range{T^{(\sym)}}|$ for every $\sym \in \alphabet$.
    Applying the union bound and adding $1$ to account for the initial state $\powerstate_\initial$ concludes the proof.
\end{proof}

Serendipitously, both the range bound and semigroup complexity can be naturally combined, admitting a more fine-grained perspective on the size of the subset automaton, each compensating for the slackness of the other.
\begin{definition}\label{def:subset-complexity}
   Given an NFA $\automaton = \fsatuple$, we define the \defn{subset complexity} \textup{$\norm{\automaton}$} as
    \begin{equation}
        \norm{\automaton} = \min_{\Gamma \subseteq \alphabet} \left(1 + \sum_{\sym \in \alphabet \setminus \Gamma} \left|\range{T^{(\sym)}} \right| \right) |\transmonoid{\Gamma}|,
    \end{equation}
    where $\Gamma$ is an arbitrary subset of the alphabet $\alphabet$ and $\transmonoid{\Gamma}$ denotes the submonoid generated by the transitions $\{ T^{(\sym)} \ | \ \sym \in \Gamma \}$.
\end{definition}
Essentially, subset complexity allows us to split $\automaton$ into subautomata according to whether the symbols on the transitions are in $\Gamma$ or $\Sigma \setminus \Gamma$, respectively.
Then, we either apply the monoid or the range bound independently.
The subset complexity is the best possible split.

\begin{theorem}[(Subset Complexity)]\label{thm:decomposition-bound}
  Let $\automaton$ be an NFA and let \textup{$\ssautomaton$} be the DFA output by the subset construction with $\automaton$ as input.
  Then, we have \textup{$|\ssautomaton| \leq \norm{\automaton}$}.
\end{theorem}
\begin{proof}
    Let $\Gamma \subseteq \alphabet$ be arbitrary. 
    We first show that
    \begin{equation}
        |\ssautomaton| \leq \left(1 + \sum_{\sym \in \Sigma \setminus \Gamma} \left|\range{T^{(\sym)}} \right| \right) |\transmonoid{\Gamma}|.
    \end{equation}
    For any word $\strx \in \kleene{\alphabet}$,
    let $\strx = \stry \strz = \stry z_1 \cdots z_m$ be the unique decomposition where $z_i \in \Gamma$ and $m \in \N$ is maximal.
 
    Furthermore, let $\powerstate_J = \trans(\initpowerstate, \strx)$.
    Similarly to the proof of Proposition \ref{prop:range}, we consider a characteristic vector $J$ for $\powerstate_J$.
    For clarity, we will denote transition matrices corresponding to symbols $\sym \in \Gamma$ with $R^{(\sym)}$ and transition matrices which correspond to symbols $\sym \notin \Gamma$ with $L^{(\sym)}$.
    Then, the characteristic vector $J$ can be written as
    \begin{equation}
        J = J' R^{(z_{1})} \cdots R^{(z_{M})},
    \end{equation}
    where $J'$ is either the initial characteristic vector $I^\top$ (in case $\strx = \strz$, i.e., $\stry = \eps$) or an element of $\bigcup_{\sym \in \alphabet \setminus \Gamma} \range{L^{(\sym)\top}}$, since the last symbol of $\stry$ lies outside of $\Gamma$.\footnote{Note that $J'$ will, in general, be the product of both $L$ and $R$ matrices.}
    Because $|\range{L^{(\sym)\top}}| = |\range{L^{(\sym)}}|$, there are at most $1 + \sum_{\sym \in \alphabet \setminus \Gamma} \left|\range{L^{(\sym)}} \right|$ choices for a vector $J'$ and $|\transmonoid{\Gamma}|$ choices for the matrix $R^{(z_{1})} \cdots R^{(z_{M})}$.
    Thus, the total number of choices for $J$ is bounded by the corresponding product. The minimum bound follows.
\end{proof}

Observe that the new notion of subset complexity is a direct improvement over both the monoid and the range bounds, since it allows us to cherry-pick individual transition matrices according to their properties to achieve the tightest bound.

\begin{figure}[tb]
    \centering
    \begin{tikzpicture}[->,>=stealth',shorten >=0.5pt,auto,node distance=15mm, scale=0.8, transform shape]
        \tikzstyle{every state}=[fill=none,draw=black,text=black,minimum size=0pt,inner sep=2.5pt]
            \node[state, initial] (q1) {$q_1$};
            \node[state, right of=q1] (q2) {$q_2$};
            \node[state, right of=q2] (q3) {$q_3$};
            \node[fill=white, right of=q3] (q?) {$\ldots$};
            \node[state, accepting, right of=q?] (qn) {$q_n$};
            \draw (q1) edge[loop above] node{$b$} (q1)
                  (q1) edge[above] node{$a$} (q2)
                  (q2) edge[above] node{$a, b$} (q3)
                  (q3) edge[above] node{$a, b$} (q?)
                  (q?) edge[above] node{$a, b$} (qn)
                  (qn) edge[bend right=1.7cm, above] node{$c$} (q2)
                  (qn) edge[bend left=1.5cm, below] node{$c$} (q1);
    \end{tikzpicture}
    \hfill
    \begin{tikzpicture}[->,>=stealth',shorten >=0.5pt,auto,node distance=15mm, scale=0.8, transform shape]
        \tikzstyle{every state}=[fill=none,draw=black,text=black,minimum size=0pt,inner sep=2.5pt]
            \node[state, accepting, initial] (q1) {$q_1$};
            \node[state, above right = {12mm} and {3mm} of q1] (q2) {$q_2$};
            \node[state, above right of=q2] (q3) {$q_3$};
            \node[state, right of=q3] (q4) {$q_4$};
            \node[fill=white, below right of=q4] (q?) {$\ldots$};
            \node[state, right = {43mm} of q1] (qn) {$q_n$};

            \draw (q1) edge[in=85,out=115,loop, above] node{$a$} (q1)
                  (q2) edge[loop above] node{$a$} (q2)
                  (q3) edge[loop above] node{$b$} (q3)
                  (q4) edge[loop above] node{$b$} (q4)
                  (qn) edge[in=50,out=80,loop, above] node{$a$} (qn)
                  (q1) edge[above left] node{$a$} (q2)
                  (q2) edge[above left] node{$a$} (q3)
                  (q3) edge[above] node{$a$} (q4)
                  (q4) edge[above right] node{$a$} (q?)
                  (q?) edge[above right] node{$a$} (qn)
                  (qn) edge[above] node{$a, b$} (q1)
                  (q2) edge[bend left=1cm,left] node{$b$} (q1)
                  (q3) edge[bend left=1cm,left] node{$b$} (q1)
                  (q4) edge[bend left=1cm,above] node{$b$} (q1);
        \end{tikzpicture}
    \caption{An $n$-state modified Moore automaton $\mooreautomatonprime$ over extended alphabet $\alphabet = \{ \syma, \symb, \symc \}$ (left) and an $n$-state modified Meyer--Fischer automaton $\mfautomatonprime$ where some labels on self loops with $\symb$ were swapped for $\syma$ (right).}
    \label{fig:modified}
\end{figure}

\begin{example}\label{ex:modified-moore}
    Consider the $n$-state Moore automaton $\mooreautomaton$ from \cref{fig:Moore-NFA} with state complexity of $2^n$.
    We define a modified $n$-state Moore automaton $\mooreautomatonprime$ as follows.
    Take the $n$-state Moore automaton and replace the labels on the transitions $\edgenoweight{\stateq_n}{a}{\stateq_1}$ and $\edgenoweight{\stateq_n}{a}{\stateq_2}$ with a new symbol $\symc \notin \{a, b\}$. The new automaton $\mooreautomatonprime$ operates over the alphabet $\alphabet = \{\syma, \symb, \symc \}$; it is still nondeterministic, as shown in \cref{fig:modified} (left).
    \begin{nclaim2}
        For an $n$-state modified Moore automaton \textup{$\mooreautomatonprime$}, the following inequality holds \textup{$\norm{\mooreautomatonprime} \leq 3n^2 + 3n$}.
    \end{nclaim2}
    Indeed, let $\Gamma = \{\syma, \symb\}$ and observe that $T^{(\syma)} T^{(\symb)} = T^{(\syma)}T^{(\syma)}$, which means that all matrices in the transition monoid $\transmonoid{\Gamma}$ can be obtained from a product sequence of the following form
    \begin{equation*}
        T^{(\symb)}T^{(\symb)}\cdots T^{(\symb)}T^{(\syma)}T^{(\syma)}\cdots T^{(\syma)}.
    \end{equation*}
    Moreover, we have that ${(T^{(\syma)})}^n = 0$, i.e., the zero matrix and ${(T^{(\symb)})}^{n-1} = {(T^{(\symb)})}^n$, meaning that there are at most $n + 1$ distinct powers of $T^{(\syma)}$ and $n$ distinct powers of $T^{(\symb)}$ (counting the empty set). 
    Hence, $|\transmonoid{\Gamma}| \leq n(n+1) = n^2 + n$.
    Moreover, we have that $|\range{T^{(\symc)}}| = 2$ because $T^{(\symc)}$ is a rank-$1$ matrix. 
    Applying \cref{thm:decomposition-bound} concludes the claim. 
\end{example}

\begin{example}\label{ex:modified-mf}
    Let us consider the modified Meyer--Fischer automaton $\mfautomatonprime$ where we replace the labels on all but $k$ of the $\symb$-labeled self-loop transitions with $\syma$, as shown in \cref{fig:modified} (right).

    \begin{nclaim3}
        For an $n$-state modified Meyer--Fischer automaton \textup{$\mfautomatonprime$} it holds: \textup{$\norm{\mfautomatonprime} \leq (2^{k+1} + 1)(n^2 - 2n + 3)$}.
    \end{nclaim3}

    To demonstrate the claim we choose $\Gamma = \{ \syma \}$ and observe that the cyclicity of $\graph(T^{(\syma)})$ is $1$ because it is strongly connected and has a self-loop. By applying Proposition \ref{prop:monoid-one-letter} we therefore obtain $|\transmonoid{\Gamma}| \leq n^2 - 2n + 3$. Finally, we notice that $|\range{T^{(\symb)}}|$ is at most $2^{k+1}$ since $T^{(\symb)}$ consists precisely of $k+1$ non-zero columns. 
    Applying \cref{thm:decomposition-bound} concludes the claim. 
\end{example}

We note that in Examples \ref{ex:modified-moore} and \ref{ex:modified-mf} both the initial states $\initial$ and the final states $\final$ can be chosen arbitrarily: the subset complexity is fully agnostic about them.

\section{Practical Considerations}\label{sec:practical}

In this section, we discuss some computationally feasible approaches for estimating the subset complexity from \cref{thm:decomposition-bound}, which requires computing the size of the range space as well as the size of the transition monoid. Computing the cardinality of the range of a Boolean matrix is equivalent to counting maximal independent sets in bipartite graphs \cite{rangeduality}, which is a $\#\pclass$-complete problem \cite{countingmaximalis}. The monoid size was partially examined in \cref{sec:semigroup-approach} and is generally difficult to determine.

However, despite these obstacles, we managed to derive a somewhat efficiently computable upper bound for subset complexity by restricting the monoid to the one-letter case only. This idea gives insight into the determinization complexity of many non-trivial NFAs, and in particular, two-letter automata.

\begin{lemma}[(Lemma 1 in \cite{boolean-width})]\label{lem:range-bounds}
    Given an $n \times n$ Boolean matrix $A$, let $\rank{A}$ be the rank of $A$ over the Galois field $\field_2$. Then the range of $A$ satisfies
    \begin{equation}
        \rank{A} \leq |\range{A}| \leq 2^{\rank{A}^2/4 + \bigO{\rank{A}}}.
    \end{equation}
\end{lemma}

\begin{remark}
    The rank over $\field_2$ can be calculated using Gaussian elimination over $\field_2$. Other possibilities for bounding the range exist. For example, one can consider the size of maximum matching or the notion of Boolean width; see \cite{boolean-width} for in-depth discussion.
\end{remark}

In \cref{sec:semigroup-approach} we have discussed the connection between the cyclicity of the precedence graph and the semigroup complexity in the case of unary automata. Using the simplified bounds for range on symbols $\alphabet \setminus \{ \syma \}$ and monoid size on the remaining symbols $\Gamma = \{ \syma \}$ we get a simple upper bound on the subset complexity for any automaton $\automaton$. 
\begin{proposition}[(All-but-one)]\label{prop:all-but-one}
    Given an $n$-state NFA $\automaton = \fsatuple$, an arbitrary target symbol $\syma \in \alphabet$ and any $\varepsilon > 0$, the following holds
    \begin{equation}
        \norm{\automaton} \in \bigO{\nsymbols(c(\graph(T^{(\syma)})) + n^2)  \max_{\symb \in \alphabet \setminus \{ \syma \}}2^{\rank{T^{(\symb)}}^2/(4 - \varepsilon)}}.
    \end{equation}
\end{proposition}

The proof is a straightforward application of Lemma \ref{lem:range-bounds} and Proposition \ref{prop:monoid-one-letter}, and gives us the following intuition about determinization: if all but one of the transition matrices are of low rank, then the subset complexity will be polynomially bounded (the cyclicity is large only in very artificial cases). 

\section{Conclusion}

Given an NFA, we have attempted to develop a method for measuring its determinization complexity, i.e., the difficulty of transforming it into the corresponding minimal DFA. First, we began by examining the shortcomings of classical complexity measures and concluded that the ideal measure should be intrinsically connected to the underlying determinization algorithms. Unfortunately, it turned out to be $\pspace$-hard to provide any accurate approximation for the performance of many major determinization algorithms under any reasonable definition thereof.

In the second part, we focused primarily on the state complexity of the subset automaton $\ssautomaton$ which, in our opinion, is integral to understanding determinization. We then linked the subset automaton with the size of the underlying transition monoid and investigated different methods to bound it. Finally, we defined a new metric---the subset complexity---and examined some subclasses of automata for which it is particularly suitable to apply.

\section*{Acknowledgments}
The authors thank the anonymous referees for several suggestions and, in particular, Ian McQuillan for fruitful discussions about generalizing results from \cref{sec:inapproximability}.

%
%
%
\biblio{jalc}\pagebreak

\begin{appendix}

\end{appendix}

\end{document}